\documentclass[11pt]{article}
\usepackage{mathrsfs}
\usepackage{amsmath}
\usepackage{amsfonts}
\usepackage{amssymb, amsmath, cite}

\newtheorem{theorem}{Theorem}
\newtheorem{lemma}{Lemma}
\newtheorem{proposition}{Proposition}
\newtheorem{remark}{Remark}

\newtheorem{corollary}{Corollary}

\newcommand\be{\begin{equation}}
\newcommand\ee{\end{equation}}
\newcommand\ber{\begin{eqnarray}}
\newcommand\eer{\end{eqnarray}}
\newcommand\berr{\begin{eqnarray*}}
\newcommand\eerr{\end{eqnarray*}}

 \newcommand\re{\mathrm{e}}
  \newcommand\ri{\mathrm{i}}
\newcommand{\ud}{\mathrm{d}}
\newcommand{\nm}{\nonumber}

\newcommand{\ito}{\int_{\Omega}}

\newcommand{\vep}{\varepsilon}

\title{\bf  Doubly Periodic  Self-Dual  Vortices for a Relativistic Non-Abelian Chern--Simons   Model}
\author{ Xiaosen Han \\Institute of Contemporary Mathematics, School  of Mathematics\\ Henan University, Kaifeng, Henan 475004, PR China
\\Gabriella Tarantello\\Dipartimento di Matematica, Unversit\`{a} di Roma ``Tor Vergata"\\ Via della Ricerca Scientifica, 00133 Rome, Italy}
\date{}

\begin{document}
\maketitle
\begin{quote}{

{{\bfseries Abstract.}
 In this paper  we establish a multiplicity  result concerning the existence of  doubly periodic  solutions for a $2\times2$ nonlinear elliptic system
 arising in the study of  self-dual non-Abelian Chern--Simons vortices. We show that the given system admits at least two solutions when the Chern--Simons coupling parameter $\kappa>0$ is
 sufficiently small;  while no solutions exist for $\kappa>0$ sufficiently large. As in  \cite{nota}, we use a variational formulation of the problem.
  Thus, we obtain a first solution via a (local) minimization method  and show that it is asymptotically gauge-equivalent to the (broken) principal embedding vacuum of the system,  as $\kappa\to 0$.
 Then we obtain the second solution by a min-max procedure of ``mountain pass'' type.  }}

\end{quote}

\section{Introduction}
\setcounter{equation}{0}

As well known vortices play an important role in many areas of
physics, including superconductivity \cite{abri,jata,gila},
optics\cite{bec}, cosmology\cite{hiki,kibb,vish}, the quantum Hall
effect \cite{soko}, and quark confinement
\cite{mayu,man1,man2,hoo1,hoo2}. After the pioneer work of
Bogomol'nyi\cite{bogo} and Prasad--Sommerfield \cite{ps},  rigorous
mathematical  results  about the existence of vortices have been
pursued in various self-dual gauge field theories on the basis of
an analytical approach that Taubes  introduced in \cite{taub1} to
treat  the Abelian--Higgs model. Indeed following \cite{taub1}, one
is  able   to reduce the vortex problem  to second order elliptic
equations with exponential nonlinearity  and Dirac source terms.
Within this framework  we mention for example the
$(2+1)$-dimensional abelian Chern--Simons model  of
Hong--Kim--Pac\cite{hkp} and Jackiw--Weinberg\cite{jw1},  for which
Taubes'  approach has  lead to the existence of  topological
multivortices (as described in  \cite{wangr,spya4}),
non-topological multivortics (as  constructed in
\cite{spya3,chmy,chim,cfl,ckl}) and  doubly periodic vortices (as
given in \cite{caya1,taran96,cho1,djlw,linyan,tara5,nota1}). In the
non-Abelian context,  rigorous existence results are established  in
\cite{sy6,sy7,chl1,bata1}, while  a series of sharp  existence
results have been obtained in  \cite{chya1,liey1, liny1,
liny2,tara1} for   non-Abelian models proposed in connection
 with the  quark confinement phenomenon  \cite{man1,man2,hoo1,hoo2}. For more results about self-dual vortices, we refer the readers to the monographs \cite{yang1, taranb}.

Here, we are going to analyze a relativistic (self-dual) non-Abelian
Chern--Simons model proposed by Dunne in  \cite{dun1,dun2}. For this
model, Yang \cite{y97} first established the existence of
topological solutions in a very general  situation. Subsequently,
for the gauge group $SU(3)$,  Nolasco and Tarantello \cite{nota}
proved a multiplicity result about the existence of doubly periodic
vortices.   The  purpose of this paper is to establish analogous multiplicity results for  theories that
involve more general  gauge groups.  More precisely, we focus on
gauge groups  with a semi-simple Lie algebra  of rank 2.

From the technical point of view, we need  to handle a $2\times2$
nonlinear elliptic system on the flat $2$--torus, with coupling
matrix given by the Cartan matrix associated to the gauge group.
Clearly, this more general situation poses new analytical
difficulties compared to the (already nontrivial) case analyzed in  \cite{nota}, where the authors handle a (specific) symmetric $2\times 2$ system. Actually, we manage to resolve such difficulties for a larger class of $2\times 2$ systems,
 where our vortex problem is included as a particular case.

\section{ Derivation of a general $2\times2$ nonlinear elliptic system and statement of the main results}
\setcounter{equation}{0}

The non-Abelian Chern--Simons model introduced by Dunne in
\cite{dun1,dun2}, is formulated over the
$\mathbb{R}^{1+2}$-Minkowski space with metric tensor:
$\mathrm{diag}(1, -1, -1)$,  that will be used  in the usual way to
rise  and lower indices. Using the summation convention over
repeated lower and upper indices ( ranging over  $0, 1, 2$),  we
consider the Lagrangian density:
 \ber
  \mathcal {L}= -\frac{\kappa}{2}{\rm Tr}\epsilon^{\mu\nu\alpha}\left(\partial_\mu A_\nu A_\alpha +\frac23A_\mu A_\nu A_\alpha\right)
  + {\rm Tr}\left([D_\mu\phi]^\dagger [D^\mu\phi]\right)-V(\phi, \phi^\dagger),\label{01}
 \eer
where $D_\mu=\partial_\mu+[A_\mu, \cdot]$ is the gauge-covariant
derivative applied to the Higgs field $\phi$ in the adjoint
representation of the gauge group $G$.  The  associated semi-simple
Lie algebra is denoted by $\mathcal {G}$, with $[\cdot, \cdot]$ the
corresponding  Lie bracket.  Moreover, $(A_\mu)_{\mu=0,1,2}$
denotes the $\mathcal{G}$-valued  gauge fields and  $\mathrm{Tr}$
refers to the trace in the matrix representation of  $\mathcal{G}$.
As usual, we denote by  $\kappa>0$ the Chern--Simons coupling
parameter, $\epsilon^{\mu\nu\alpha}$ the Levi--Civita totally
skew-symmetric tensor with $\vep^{012}=1$ and we let  $V$  be the Higgs
potential.

The Euler-Lagrange equations corresponding to  \eqref{01} are given
by
  \ber
   D_\mu D^\mu \phi&=&\-\frac{\partial V}{\partial \phi^\dagger},\label{02}\\
    \kappa F_{\mu\nu }&=&\epsilon_{\mu\nu\alpha} J^{\alpha},\label{03}
  \eer
with the strength tensor:
 \be
   F_{\mu\nu}=\partial_\mu A_\nu-\partial_{\nu}A_\mu+[A_\mu, A_\nu],\label{04}
 \ee
and  covariant current density:
 \be
  J^\mu=[\phi^\dagger, (D^\mu\phi)]-[(D^\mu\phi)^\dagger, \phi],\label{05}
  \ee
which is conserved, by satisfying:
  \[D_\mu J^\mu=0.\]
 The system also admits a conserved Abelian current:
  \[ Q^\mu=-\ri\mathrm{Tr}\left(\phi^\dagger D^\mu\phi-(D^\mu\phi)^\dagger\phi\right), \quad \mu=0, 1, 2;\]
 that satisfies: $\partial_\mu Q^\mu=0$,   and it is  due to the  global $U(1)$-invariance of the system.

Note that the energy density associated  to \eqref{01}  is given by:
 \be
  \mathcal {E}={\rm Tr}([D_0\phi]^\dagger[D_0\phi])+{\rm Tr}([D_i\phi]^\dagger[D_i\phi])+V(\phi, \phi^\dagger), \label{06}
  \ee
 that we consider together  with the following Gauss law of the system:
  \ber
   \kappa F_{12}=J^0=[\phi^\dagger, (D_0\phi)]-[(D_0\phi)^\dagger, \phi]\label{07}
  \eer
(corresponding to the $\alpha=0$ component of \eqref{03}).  Then, with the
choice of  the  Higgs potential:
   \[V(\phi, \phi^\dagger)=\frac{1}{\kappa^2} {\rm Tr}\left\{\left([[\phi,\phi^\dagger], \phi]-v^2\phi\right)^\dagger\left([[\phi,\phi^\dagger], \phi]-v^2\phi\right)\right\}, \]
 ( $v>0$ is a constant which measures the scale of the broken symmetry)  we see that the energy density $\mathcal {E}$ can be shown to satisfy (\cite{dun1,dun2,yang1})
    \[ \mathcal{E}\ge \frac{v^2}{\kappa}Q_0 \]
(neglecting divergence terms). Moreover, the above lower bound is
saturated by field configurations satisfying the following
relativistic Chern--Simons
 self-dual equations:
   \ber
    D_1\phi\pm\ri D_2\phi&=&0, \label{09}\\
    \ri F_{12}\mp\frac{2}{\kappa^2}[[[\phi,\phi^\dagger], \phi]-v^2\phi, \phi^\dagger]&=&0.\label{010}
   \eer
See \cite{dun1,dun2,yang1} for details. It is not difficult to see
that  the solutions of  \eqref{09} and \eqref{010} also satisfy the
Euler--Lagrange equations \eqref{02} and \eqref{03}.

To  handle  the self-dual equations \eqref{09} and \eqref{010}, we
follow \cite{dun1}, and use the following decomposition:
 \ber
  A_\mu=\ri\sum\limits_{j=1}^r A_\mu^jH_j, \quad
    \phi=\sum\limits_{i=1}^r\phi^iE_i, \label{011}
 \eer
where  $A_\mu^i$ are real-valued vector fields, $\phi^i$ are
complex valued scalar fields $(i=1, \dots, r)$,  $r$ is the rank of
the semi-simple Lie algebra $\mathcal{G}$, $\{H_i\}_{1\le i\le r}$
and $\{E_i\}_{1\le i\le r}$ (with $E_i^\dagger=E_{-i}$) are the
generators of the Cartan subalgebra and the family of simple  ladder
operators of the semi-simple Lie algebra $\mathcal{G}$,
respectively.  The consistency of \eqref{011} can be checked on the
basis of  the following commutation and trace relation,
 \berr
  [H_i, H_j]&=&0, \\
  \left[E_i, E_{-j}\right]&=&\delta_{ij}H_i,\\
   \left[H_i, E_{\pm j}\right]&=&\pm K_{ij}E_{\pm j},\\
    {\rm Tr}(E_iE_{-j})&=&\delta_{ij},\\
    {\rm Tr}(H_iH_j)&=&K_{ij},\\
      {\rm Tr}(H_iE_{\pm j})&=&0, \quad i,j=1,\dots, r,
 \eerr
where $K=(K_{ij})_{i,j=1,\dots, r}$ is the Cartan matrix \cite{grm}
of the semi-simple Lie algebra $\mathcal{G}$. It is  well-known that the
entries
 $K_{ij}$ of the Cartan matrix $K$, satisfy the following properties:
   \begin{quote}{
    i) if $i=j\in\{1,\dots, r\}$  then $K_{jj}=2$,

    ii) If $i\neq j\in \{1, \dots, r\}$ then  $K_{ij}\in \mathbb{Z}^-$ and $K_{ij}=0\Leftrightarrow K_{ji}=0$. }
    \end{quote}
We also know that for a semisimple Lie algebra,
  \be
  \det K>0, \label{012}
  \ee
 (in fact all its principal  diagonal minors are positive), and so $K$ is  non-degenerate.  Actually, i) and ii) also imply that the entries
 of the inverse matrix $K^{-1}$ are all non-negative, see \cite{grm} for details.  Going back to \eqref{011}, we observe that it always admits a
 (trivial) zero-energy configuration for which  all the gauge fields vanish,  while the Higgs field $\phi$ satisfies:
   \be
  [[\phi,\phi^\dagger], \phi]-v^2\phi=0. \label{013}
  \ee
 All such vacua configurations correspond to minima for the given potential.

 In particular, using the decomposition \eqref{011}, we can identify  the so-called principal embedding vacuum:
 $\phi_{(0)}=\sum\limits_{j=1}^r\phi^j_{(0)}E_j$  whose components $\phi^j_{(0)}$ satisfy:
   \be
  \left|\phi_{(0)}^i\right|^2=v^2\sum\limits_{j=1}^r(K^{-1})_{ij}, \quad i=1,
  \dots, r. \label{014}
  \ee

To obtain non-trivial (self-dual) vortex configurations, we use the
following standard  notations  \cite{dun1,yang1}:
 \[\partial_{\pm}=\partial_1\pm\ri\partial_2,\quad A^i_{\pm}=A_1^i\pm\ri A_2^i,  \quad  i=1,2\]  and observe that,
 in the static case, the  self-dual equations \eqref{09}-\eqref{010}   can be expressed  componentwise as follows:
 \ber
 \partial_{\pm}\ln \phi^a&=&-\ri \sum\limits_{b=1}^rA_{\pm}^bK_{ba}, \label{f1}\\
 F_{12}^a&=&\pm\frac{2}{\kappa^2}\left(\sum\limits_{b=1}^r|\phi^a|^2|\phi^b|^2K_{ba}-v^2|\phi^a|^2\right), \label{f2}
 \eer
away from the zeros of $\phi^a$,   and with
\[ F_{12}^a=\partial_1A_2^a-\partial_2A_1^a, \quad a=1, \dots, r.\]
 Following \cite{taub1}, we can combine  equations  \eqref{f1}-\eqref{f2} into  the following $r \times r$ system (so called Master equations):
 \ber
 \Delta\ln|\phi^a|^2=\pm2\sum\limits_{b=1}^rF^b_{12}K_{ba}
 =\frac{4}{\kappa^2}\left(\sum\limits_{b=1}^r\sum\limits_{c=1}^r|\phi^b|^2|\phi^c|^2K_{cb}K_{ba}-v^2\sum\limits_{b=1}^r|\phi^b|^2K_{ba}\right), \label{f3}
 \eer
(away from the zero points of $\phi^a$)   $a=1,\dots, r$,   that we need to solve in combination with the following componentwise expression  of
the Gauss law \eqref{07}:
  \be
  \kappa F^a_{12}=J^a_0, \quad a=1, \dots, r,\label{015}
  \ee
with $J^a_0$ the component relative to the Cartan subalgebra of the
current $J_0$ in \eqref{05}.

The corresponding energy density takes the form:
 \be
   \mathcal{E}=v^2\sum\limits_{a=1}^r F^a_{12}.\label{016}
  \ee
 While, the gauge invariance of the theory  is expressed by the following transformation laws:
  \be
   A^a_\mu\rightarrow A^a_\mu+\partial_\mu \omega_a \quad  \mu=0, 1, 2,\quad \phi^a\rightarrow
   \re^{\ri\sum\limits_{b=1}^rK_{ba}\omega_b}\phi^a\label{017}
  \ee
with $\omega_a$ a smooth real function,  that in the static case
depends only on the state variables $x=(x_1, x_2)\in \mathbb{R}^2,
\, a=1, \dots, r$.

In this paper,  we are interested  in  obtaining static solutions of
\eqref{f3} subject to  suitable 't Hooft boundary conditions over a
doubly periodic domain $\Omega$. To be more precise, we let the
periodic cell domain  $\Omega$ to be generated by two linearly
independently vectors $\mathbf{e}_1, \mathbf{e}_2\in \mathbb{R}^2$,

  \[ \Omega=\left\{x=s_1\mathbf{e}_1+s_2\mathbf{e}_2\in \mathbb{R}^2\quad  0<s_j<1, \quad j=1,2\right\}, \]
and  set
 \[ \Gamma_j=\big\{x=s_j\mathbf{e}_j, \quad 0<s_j<1\big\}, \quad j=1, 2\]
 so that
   \[\partial\Omega=\Gamma_1\cup\Gamma_2\cup\{\mathbf{e}_1+\Gamma_2\}\cup\{\mathbf{e}_2+\Gamma_1\}\cup\{0, \mathbf{e}_1, \mathbf{e}_2
   , \mathbf{e}_1+\mathbf{e}_2\}.\]

 In view of \eqref{017}, we require  $(A^a_\mu)_{\mu=0,1,2}$ and  $\phi^a$ to satisfy the boundary conditions
   \ber\left\{\begin{array}{lll}
   \left(\re^{\ri\sum\limits_{b=1}^rK_{ba}\omega_b}\phi^a\right)(x+\mathbf{e}_k)=\left(\re^{\ri\sum\limits_{b=1}^rK_{ba}\omega_b}\phi^a\right)(x),\\[3mm]
    \left(A_\mu^a+\partial_\mu\omega_k^a\right)(x+e_k)=\left(A_\mu^a+\partial_\mu\omega_k^a\right)(x), \quad \mu=0, 1,
    2,\\[3mm]
    x\in \Gamma_1\cup\Gamma_2\setminus\Gamma_k, \quad k=1, 2,\quad   a=1, \dots, r,
    \end{array}
    \right. \label{018}
   \eer
where  $\omega_k^a$ is a smooth function defined in  a neighborhood
of $\Gamma_j\cup\{\Gamma+\mathbf{e}_k\}$ with $j\neq k\in\{1, 2\},
\, a=1, \dots, r$.

As explicitly derived in \cite{nota}, solutions of \eqref{f3} and
\eqref{018}  carry ``quantized'' electric and magnetic charges, in
the sense that the following hold:
  \ber
   \Phi_a:&=&\int F^a_{12}= 2\pi \sum\limits_{b=1}^r(K^{-1})_{ba}N_b, \label{018a}\\
    Q_a:&=&\int J^0=\kappa\Phi_a=2\pi\kappa\sum\limits_{b=1}^r(K^{-1})_{ba}N_b, \quad a=1, \dots, r\label{018b}
  \eer
with $N_a$ a suitable integer, that actually  counts the zeros of
$\phi^a$ in $\Omega$ (with  multiplicity)  $a=1, \dots, r$.

 In addition, from \eqref{016}, \eqref{018a}  and \eqref{018b}, we obtain  the following ``quantization'' formula for the total energy:
    \be
     E=\ito\mathcal{E}=2\pi v^2\sum\limits_{a, b=1}^r(K^{-1})_{ab}N_b=2\pi \sum\limits_{b=1}^r|\phi_{(0)}^b|^2N_b,\label{018c}
    \ee
where the last identity follows by \eqref{014}, with $\phi_{(0)}^b$
the component of the principal embedding vacuum.

Here we shall focus on the solvability  of \eqref{f3} and
\eqref{018}  with gauge groups rank $r=2$.  Besides the group $SU(3)$,
with Cartan  matrix
   $K=\begin{pmatrix}
2 & -1 \\
-1 & 2 \\
\end{pmatrix},$ examples of this situation include the exceptional  gauge group $B_2 (=C_2)$ with Cartan matrix
$ K=\begin{pmatrix}
2 & -1 \\
-2 & 2 \\
\end{pmatrix}
$
 and  $G_2$ with Cartan matrix
 $
K=\begin{pmatrix}
2 & -1 \\
-3 & 2 \\
\end{pmatrix}.
$

More generally,  in the rank $r=2$ case, the Cartan matrix takes the
form
 \be
K=\begin{pmatrix}
2 & -a_{12} \\
-a_{21} & 2 \\
\end{pmatrix} \label{018d}
\ee
 with $a_{jk}\in\mathbb{Z}^+$ for $j\neq k\in \{1, 2\}$ and $4-a_{12}a_{21}>0$.

In case $a_{12}=0=a_{21}$ (i.e. $G=A_1\times A_1$) then the Cartan
matrix diagonalizes, and the system \eqref{f3} decouples into two
abelian Chern--Simons vortex problems, for which the existence of
(at least) two gauge-distinct periodic static configurations has
been established in \cite{taran96}, provided  $\kappa>0$ is
sufficiently small. Our main goal is to extend such multiplicity
result to any gauge group of rank $2$.  More precisely, we prove:
 \begin{theorem} \label{tha1}
Let the gauge group $G$ admit a semisimple Lie algebra $\mathcal{G}$
of rank $r=2$ and Cartan matrix $K$ specified  in \eqref{018d}.  For
$N_a\in \mathbb{N}$, let $Z_a=\{p_{a,1}, \dots,
p_{a,N_a}\}\subset\Omega$ be a set of $N_a$ points (not necessarily
distinct) $a=1, 2$.  For $\kappa>0$ sufficiently small,  there exist
at least  two  gauge distinct static solutions of
\eqref{09}-\eqref{010} subject to the ansatz \eqref{011} and the
boundary condition \eqref{018} such that:

(i) the component $\phi^a$ of the Higgs field satisfies:
$|\phi^a|<|\phi_{(0)}^a|$ in $\Omega$, with $\phi_{(0)}^a$ the
  component of the principal embedding vacuum in \eqref{014}; and   $\phi^a$ vanishes exactly at each point $p_{a,j}\in Z_a$ with  the same multiplicity, $a=1, 2$;

(ii) the corresponding magnetic flux $\Phi_a$, electric charge
$Q_a(a=1, 2)$  and total energy $E$, satisfy the ``quantization''
identity \eqref{018a}, \eqref{018b} and \eqref{018c} respectively;

(iii)  for at least one of  the given solutions  the following
holds:
      \[ |\phi^a|\to|\phi_{(0)}^a|\quad \text{as}\quad \kappa\to 0, \]
 pointwise a.e. in $\Omega$ and strongly in $L^p(\Omega)$, for $p\ge1$.

(iv) If
 \[\kappa>v \sqrt{\frac{|\Omega|}{4\pi (\det K)\max\left\{\frac{2N_1+a_{12}N_2}{2(a_{12}+2)^2},\frac{2N_2+a_{21}N_1}{2(a_{21}+2)^2}\right\}}},\]
with $a_{ij}\ge0,   \, i\neq j\in \{1, 2\} $, the off-diagonal
entries of the Cartan matrix  $K$ in \eqref{018d}, then problem \eqref{09}-\eqref{011} and
\eqref{018} admits no such solutions .
\end{theorem}

As already mentioned,  Theorem \ref{tha1} provides a natural
extension of the multiplicity result of Nolasco--Tarantello in
\cite{nota}, concerning the group $G=SU(3)$, for which \eqref{f3}
enjoys additional symmetries.  In fact, to establish Theorem
\ref{tha1} we adopt the same variational viewpoint. However we are
able to   handle systems of the type \eqref{f3} with a more general
coupling matrix.

More precisely, we take a $2\times2$ matrix $K$ of the form:
\begin{equation*}
K=\begin{pmatrix}
a & -b \\
-c & d \\
\end{pmatrix},
\end{equation*}
 and assume that  $a, b, c, d>0$ and   $ad-bc>0$. Notice that the case  $a, d>0$ and $b=c=0$, is already covered in \cite{taran96}.

 We denote the zero set  of $\phi^i$ by
 \ber
 Z_i=\{p_{i,1}, \dots, p_{i, N_i}\}, \quad\,\, i=1, 2\label{2.10}
 \eer
 (repeated with multiplicity) and set,
 \be
 \left|\phi^1\right|^2=v^2\frac{b+d}{ad-bc}\re^{u_1}, \quad
 \left|\phi^2\right|^2=v^2\frac{a+c}{ad-bc}\re^{u_2},\quad
 \lambda=\frac{4v^4}{\kappa^2}. \label{f5}
 \ee

By straightforward calculations,  we see that 
the equations in \eqref{f3}  subject to the boundary conditions
\eqref{018} take the form:
   \ber\left\{\begin{array}{lll}
    \Delta u_1&=&
    \lambda\left\{\frac{1}{ad-bc}\left[-a(b+d)\re^{u_1}+b(a+c)\re^{u_2}\right]\right. \\
    &&\left.+\frac{1}{(ad-bc)^2}\left[a^2(b+d)^2\re^{2u_1}-b(b+d)(a^2-c^2)\re^{u_1+u_2}-bd(a+c)^2\re^{2u_2}\right]\right\} \\
    &&+4\pi\sum\limits_{j=1}^{N_1}\delta_{p_{1,j}},\,x\in \Omega,\\
      \Delta u_2&=&\lambda\left\{\frac{1}{ad-bc}\left[c(b+d)\re^{u_1}-d(a+c)\re^{u_2}\right]\right. \\
    &&\left.+\frac{1}{(ad-bc)^2}\left[-ac(b+d)^2\re^{2u_1}-c(a+c)(d^2-b^2)\re^{u_1+u_2}+d^2(a+c)^2\re^{2u_2}\right]\right\} \\
    &&+4\pi\sum\limits_{j=1}^{N_2}\delta_{p_{2,j}}\,x\in \Omega,\\
     u_1 & \text{and} & u_2 \quad \text{ doubly periodic on} \quad \partial\Omega.
 \end{array}\label{a1}
\right.
  \eer

 Concerning \eqref{a1},  we establish the following:
\begin{theorem}\label{thb1}
   Assume that  $a, b, c, d>0$ and $ad-bc>0$.  Given $N_j\in
   \mathbb{N}$ and $Z_j=\{p_{j,1}, \dots, p_{j, N_j}\}\subset\Omega$ (a set of $N_j$-point
   repeated with multiplicity), $j=1, 2$, the following holds:

 1.  Every solution $(u_1, u_2)$ of
 \eqref{a1}  satisfies
   \be
    \re^{u_1}<1, \quad \re^{u_2}< 1 \quad in\quad  \Omega.\label{b02}
    \ee

2. If
   \[\lambda<\frac{16\pi(ad-bc)}{|\Omega|}\max\left\{\frac{dN_1+bN_2}{a(b+d)^2}, \frac{cN_1+aN_2}{d(a+c)^2}\right\},\]
   then   problem \eqref{a1}   admits  no solutions.

 3. There exist $\lambda_0>0$,  such that  for $\lambda>\lambda_0$  problem \eqref{a1}  admits at least   two distinct solutions, one of which satisfying:
   \be
   \re^{u_1}\to 1, \quad\re^{u_2}\to 1, \quad as \quad \lambda \to +\infty \label{b04}
   \ee
 pointwise a.e. in $\Omega$ and strongly in $L^p(\Omega)$ for any $p\ge1$.
\end{theorem}

 \begin{remark}
  As already noticed,  when   $b=c=0$,  problem   \eqref{a1}  decouples in  two  abelian self-dual Chern-Simons equations:
\berr\left\{\begin{array}{lll}
\Delta u_i=\lambda\re^{u_i}\left(\re^{u_i}-1\right)+4\pi\sum\limits_{j=1}^{N_i}\delta_{p_{i,j}},\,x\in  \Omega,\\
u_i \quad \text{doubly periodic on }\quad \partial\Omega, \quad i=1,
2
 \end{array}\right.
  \eerr
for which the existence and multiplicity results claimed above have
been established in \cite{taran96}.
\end{remark}

 Thus in view of \eqref{f5},   \cite{taran96}  and  Theorem \ref{thb1}, we deduce (by standard arguments \cite{taub1}) the statement
of Theorem \ref{tha1}. Hence we devote  the following section  to the proof of 
  Theorem \ref{thb1}.

\section{ Existence of doubly periodic solutions }
\setcounter{equation}{0}\setcounter{remark}{0}

In this section we analyze problem \eqref{a1}, and  for convenience
we  rewrite  it as follows:
 \ber\left\{\begin{array}{lll}
    \Delta u_1&=&
    \frac{\lambda}{ad-bc}\left[a(b+d)\re^{u_1}\left(\re^{u_1}-1\right)+b(a+c)\re^{u_2}\left(1-\re^{u_2}\right)\right]\\
    && +\frac{\lambda(a+c)(b+d)b}{(ad-bc)^2}\left(a\re^{u_1}+c\re^{u_2}\right)\left(\re^{u_1}-\re^{u_2}\right)+4\pi\sum\limits_{j=1}^{N_1}\delta_{p_{1,j}},\,x\in \Omega,\\
      \Delta u_2&=&
    \frac{\lambda}{ad-bc}\left[d(a+c)\re^{u_2}\left(\re^{u_2}-1\right)+c(b+d)\re^{u_1}\left(1-\re^{u_1}\right)\right]\\
    && +\frac{\lambda(a+c)(b+d)c}{(ad-bc)^2}\left(b\re^{u_1}+d\re^{u_2}\right)\left(\re^{u_2}-\re^{u_1}\right)+4\pi\sum\limits_{j=1}^{N_2}\delta_{p_{2,j}},\,x\in \Omega,\\
     u_1 & \text{and} & u_2 \quad \text{ doubly periodic on} \quad \partial\Omega.
 \end{array}\label{ab1}
\right.
  \eer

We start to establish the following:
 \begin{proposition} \label{pr1}
  Let $(u_1, u_2)$  satisfy \eqref{ab1}. Then $u_i<0$ in $\Omega$, $i=1, 2$.
  \end{proposition}

 {\bf Proof.}\quad
Notice that  $u_i$ attains  its maximum  value  at a point    $
\tilde{x}_i\in \overline{\Omega}\setminus Z_i$, so that
$\tilde{u}_i\equiv\max\limits_{\overline{\Omega}}u_i=u_i(\tilde{x}_i),
\,i=1, 2$. We start by showing  $\tilde{u}_i\le 0$, for $i=1, 2$.
Indeed, in case  $\tilde{u}_1\ge \tilde{u}_2$, then we use the first
equation in \eqref{ab1} to obtain:
   \berr
   0\ge\Delta u_1(\tilde{x}_1)&=&\frac{\lambda}{ad-bc}\left[a(b+d)\re^{\tilde{u}_1}\left(\re^{\tilde{u}_1}-1\right)+b(a+c)\re^{u_2(\tilde{x}_1)}\left(1-\re^{u_2(\tilde{x}_1)}\right)\right]\\
    && +\frac{\lambda(a+c)(b+d)b}{(ad-bc)^2}\left(a\re^{\tilde{u}_1}+c\re^{u_2(\tilde{x}_1)}\right)\left(\re^{\tilde{u}_1}-\re^{u_2(\tilde{x}_1)}\right)\\
    &\ge&\frac{\lambda}{ad-bc}\left[a(b+d)\re^{\tilde{u}_1}-b(a+c)\re^{u_2(\tilde{x}_1)}\right]\left(\re^{\tilde{u}_1}-1\right).
   \eerr
Since
  \[a(b+d)\re^{\tilde{u}_1}-b(a+c)\re^{u_2(\tilde{x}_1)}\ge(ad-bc)\re^{\tilde{u}_1} >0\]
we find that necessarily,   $\tilde{u}_1\le 0$, and the desired conclusion follows in this case. On the other hand if 
$\tilde{u}_2\ge\tilde{u}_1$, then we can use a similar argument for
the second equation in \eqref{ab1} to deduce that
 $\tilde{u}_2\le0$.  Thus, in any case, we have: $\tilde{u}_i\le 0, \, i=1, 2.$ To obtain that actually the strict inequality holds, we use the strong
 maximum principle. It can be applied, since for example  we see that $u_1$ satisfies:
   \berr
    \Delta u_1+c_1(x)u_1&=&\frac{\lambda}{ad-bc}\left[b(a+c)\re^{u_2}+\frac{(a+c)(b+d)b}{ad-bc}\left(a\re^{u_1}+c\re^{u_2}\right)\right]\left(1-\re^{u_2}\right)\\
    &\ge&0 \quad \text{in}\quad \Omega
   \eerr
with
 \berr
 c_1(x)=\frac{\lambda}{ad-bc}\left[a(b+d)\re^{u_1}+\frac{(a+c)(b+d)b}{ad-bc}\left(a\re^{u_1}+c\re^{u_2}\right)\right]\frac{1-\re^{u_1}}{u_1}.
 \eerr
 Similarly for $u_2$.

 Therefore we   conclude that $u_i< 0 $ in   $\Omega$, $ i=1, 2$.   In particular  we  have established the first conclusion of Theorem \ref{thb1}. $\square$
\\[3mm]

To proceed further, we let  $u_i=u_0^i+v_i, i=1,2$ with $u_0^i  $
being  the unique solution of the problem (see \cite{aubi})
     \berr\left\{\begin{array}{lll}
    \Delta u_0^i=4\pi\sum\limits_{s=1}^{N_i}\delta_{p_{i,  s}}-\frac{4\pi N_i}{|\Omega|},\\
      \ito u_0^i\ud x=0, \quad u_0^i \quad \text{doubly periodic on}\quad  \partial\Omega\quad i=1, 2.
     \end{array}\right.
     \eerr
 Consequently, problem \eqref{a1} (or \eqref{ab1}) can be formulated in terms of the unknown $(v_1, v_2)$ as follows:
 \ber\left\{\begin{array}{lll}
  \Delta v_1=\frac{\lambda}{ad-bc}\left[-a(b+d)\re^{u_0^1+v_1}+b(a+c)\re^{u_0^2+v_2}\right]
  +\frac{\lambda}{(ad-bc)^2}\left[a^2(b+d)^2\re^{2u_0^1+2v_1}\right.\\[3mm]\left.\quad\qquad-b(b+d)(a^2-c^2)\re^{u_0^1+u_0^2+v_1+v_2}-bd(a+c)^2\re^{2u_0^2+2v_2}\right]
   +\frac{4\pi N_1}{|\Omega|}, \\[3mm]
    \Delta v_2= \frac{\lambda}{ad-bc}\left[c(b+d)\re^{u_0^1+v_1}-d(a+c)\re^{u_0^2+v_2}\right]
  +\frac{\lambda}{(ad-bc)^2}\left[-ac(b+d)^2\re^{2u_0^1+2v_1}\right.\\[3mm]
  \left.\quad\qquad-c(a+c)(d^2-b^2)\re^{u_0^1+u_0^2+v_1+v_2}+d^2(a+c)^2\re^{2u_0^2+2v_2}\right] +\frac{4\pi N_2}{|\Omega|},\\[3mm]
    v_1, \, v_2\,\, \text{doubly periodic on}\quad \partial\Omega.
    \end{array}\right.\label{a3}
   \eer

   Actually, to emphasize the variational structure of \eqref{a3}, we shall use the following equivalent  formulation:
  \ber\left\{\begin{array}{lll}
  \frac{ad-bc}{b+d} \Delta (dv_1+bv_2)= \lambda\left[a(b+d)\re^{2u_0^1+2v_1}-(ad-bc)\re^{u_0^1+v_1}-b(a+c)\re^{u_0^1+u_0^2+v_1+v_2}\right] \\[3mm]
   \qquad\qquad\qquad\qquad\quad+\frac{4\pi(ad-bc)(dN_1+bN_2)}{(b+d)|\Omega|},\\[3mm]
  \frac{ad-bc}{a+c}\Delta (cv_1+av_2)= \lambda\left[d(a+c)\re^{2u_0^2+2v_2}-(ad-bc)\re^{u_0^2+v_2}-c(b+d)\re^{u_0^1+u_0^2+v_1+v_2}\right]\\[3mm]
 \qquad\qquad\qquad\qquad\quad +\frac{4\pi (ad-bc)(cN_1+aN_2)}{(a+c)|\Omega|},\\[3mm]
    v_1, \, v_2\, \,\text{doubly periodic on} \quad \partial\Omega.
   \end{array}\right. \label{a5}
  \eer

We introduce the Hilbert space: $H(\Omega)\equiv W^{1,
2}\left(\frac{\mathbb{R}^2}{\mathbb{Z}\mathbf{e}_1+\mathbb{Z}\mathbf{e}_2}\right)$
of $\Omega$-periodic $L^2$-functions whose  derivatives  also belong
to  $L^2(\Omega)$, equipped  with the usual  norm:
 $\|w\|^2=\|w\|_2^2+\|\nabla w\|_2^2=\ito w^2\ud x+\ito |\nabla w|^2\ud x$,  $w\in H(\Omega)$.

It is not difficult to check  that weak  solutions to \eqref{a5} are
critical points  in $H(\Omega)\times H(\Omega)$ of the functional:
 \ber
 I_\lambda(v_1, v_2)&=&\frac{d}{2b}\|\nabla v_1\|_2^2+\frac{a}{2c}\|\nabla v_2\|_2^2+\ito\nabla v_1\cdot\nabla v_2\ud x
 +\lambda\ito Q(v_1, v_2)\ud x\nm\\
  &&+ \frac{\alpha_1}{|\Omega|}\ito v_1\ud x+\frac{\alpha_2}{|\Omega|}\ito v_2\ud x,\label{a7}
 \eer
where
  \ber\left\{\begin{array}{lll}
  Q(v_1,v_2)=\frac{1}{2ab(ad-bc)}Q_1^2(v_1,v_2)+\frac{(a+c)^2}{2ac}Q_2^2(v_2),\\[3mm]
 Q_1(v_1,v_2)=\left[a(b+d)\re^{u_0^1+v_1}-b(a+c)\re^{u_0^2+v_2}-(ad-bc)\right],\\[3mm]
 Q_2(v_2)=\left(\re^{u_0^2+v_2}-1\right),\\[3mm]
 \alpha_1=4\pi\left(\frac dbN_1+N_2\right), \quad
 \alpha_2=4\pi\left(N_1+\frac ac N_2\right).
 \end{array}\right.\label{a7'}
  \eer

In view of our assumption, notice that  the quadratic part of
$I_\lambda$ is positive definite.

In fact we obtain a first critical point for $I_\lambda$ via (local)
minimization.

\subsection{Constrained minimization}

For a solution  $(v_1, v_2)$ of \eqref{a5}, after integration   over
$\Omega$, we  find  the following natural  constraints:
 \ber
  &&a(b+d)\ito\re^{2u_0^1+2v_1}\ud x-(ad-bc)\ito\re^{u_0^1+v_1}\ud x-b(a+c)\ito\re^{u_0^1+u_0^2+v_1+v_2}\ud x\nm\\
  &&+\frac{4\pi(ad-bc)(dN_1+bN_2)}{\lambda(b+d)}=0,\label{a8}\\
   && d(a+c)\ito\re^{2u_0^2+2v_2}\ud x-(ad-bc)\ito\re^{u_0^2+v_2}\ud x-c(b+d)\ito\re^{u_0^1+u_0^2+v_1+v_2}\ud x
  \nm\\
   &&+\frac{4\pi (ad-bc)(cN_1+aN_2)}{\lambda (a+c)}=0.\label{a9}
 \eer

From \eqref{a7'}-\eqref{a9}, we  obtain
 \ber
 \ito Q(v_1, v_2)\ud x&=&\frac{1}{2}\left[\left(1+\frac db\right)\ito\big(1-\re^{u^1_0+v_1}\big)\ud x+\left(1+\frac ac\right)\ito\big(1-\re^{u_0^2+v_2}\big)\ud x\right]
     \nm\\&& -\frac{\alpha_1+\alpha_2}{2\lambda}.\label{a9'}
      \eer

 Therefore, if we  decompose $v_1, v_2$ as follows:
  \[v_i=w_i+c_i, \quad \ito w_i\ud x=0, \, c_i=\frac{1}{|\Omega|}\ito v_i\ud x, \quad i=1, 2,\]
then form \eqref{a8} and \eqref{a9} we find:
  \ber
   \re^{2c_1}\ito\re^{2u_0^1+2w_1}\ud x-\re^{c_1}R_1(w_1, w_2, \re^{c_2})+\frac{4\pi(ad-bc)(dN_1+bN_2) }{\lambda a(b+d)^2}=0,\label{a10}\\
   \re^{2c_2}\ito\re^{2u_0^2+2w_2}\ud x-\re^{c_2}R_2(w_1, w_2, \re^{c_1})+\frac{4\pi(ad-bc)(cN_1+aN_2) }{\lambda d(a+c)^2}=0,\label{a11}
  \eer
with
  \ber
   R_1(w_1, w_2, \re^{c_2})&=&\frac{ad-bc}{a(b+d)}\ito\re^{u_0^1+w_1}\ud x+\frac{b(a+c)}{a(b+d)}\re^{c_2}\ito\re^{u_0^1+u_0^2+w_1+w_2}\ud  x, \\\label{a12}
   R_2(w_1, w_2, \re^{c_1})&=&\frac{ad-bc}{d(a+c)}\ito\re^{u_0^2+w_2}\ud x+\frac{c(b+d)}{d(a+c)}\re^{c_1}\ito\re^{u_0^1+u_0^2+w_1+w_2}\ud x. \label{a13}
  \eer

   A necessary condition for the solvability of  \eqref{a10} and \eqref{a11}   with
respect to $c_1 $ and $c_2$  is that,
 \ber
 (R_1(w_1, w_2,  \re^{c_2}))^2&\ge&\frac{16\pi(ad-bc)(dN_1+bN_2) }{\lambda a(b+d)^2}\ito\re^{2u^1_0+2w_1}\ud x,\label{a14}\\
  (R_2(w_1, w_2, \re^{c_1}))^2&\ge&\frac{16\pi(ad-bc)(cN_1+aN_2) }{\lambda d(a+c)^2}\ito\re^{2u^2_0+2w_2}\ud x.\label{a15}
 \eer

On the other hand, from Proposition \ref{pr1}, we see that
$u^1_0+v_1+c_1<0$ and $ u^2_0+v_2+c_2< 0$ in $ \Omega$. Therefore,
as a consequence of \eqref{a14}-\eqref{a15}, we obtain:
 \berr
  \frac{16\pi(ad-bc)(dN_1+bN_2) }{\lambda a(b+d)^2}\ito\re^{2u^1_0+2w_1}\ud x\le |\Omega|\ito\re^{2u^1_0+2w_1}\ud x,\\
  \frac{16\pi(ad-bc)(cN_1+aN_2) }{\lambda d(a+c)^2}\ito\re^{2u^2_0+2w_2}\ud x\le |\Omega|\ito\re^{2u^2_0+2w_2}\ud x.
 \eerr
 Thus,  we obtain the following necessary condition for the solvability of \eqref{a5},
 \be
  \lambda\ge \frac{16\pi(ad-bc)}{|\Omega|}\max\left\{\frac{dN_1+bN_2}{a(b+d)^2}, \frac{cN_1+aN_2}{d(a+c)^2}\right\}, \label{a16}
 \ee
 and deduce  part 2.  of  Theorem \ref{thb1}.

 Conditions \eqref{a14} and \eqref{a15} suggest to focus only with   pairs $(v_1, v_2)$  that,  under the  decomposition: $v_i=w_i+c_i, \, i=1, 2$,  satisfy:
 \ber
  \ito w_1\ud x=0\quad \text{and}\quad\left(\ito\re^{u^1_0+w_1}\ud x\right)^2&\ge&\frac{16\pi a(dN_1+bN_2)}{\lambda(ad-bc)}\ito\re^{2u^1_0+2w_1}\ud x,\label{a17}\\
     \ito w_2\ud x=0\quad \text{and}\quad\left(\ito\re^{u^2_0+w_2}\ud x\right)^2&\ge&\frac{16\pi d(cN_1+aN_2)}{\lambda(ad-bc)}\ito\re^{2u^2_0+2w_2}\ud x;\label{a18}
 \eer
  and where $(c_1, c_2)$ satisfy \eqref{a10} and \eqref{a11}.

   Hence we define the admissible set:
  \be
  \mathcal {A}=\big\{(w_1, w_2)\in H(\Omega)\times H(\Omega)\quad \text{such that } \quad\eqref{a17}\quad \text{and} \quad\eqref{a18}\,\,\text{hold} \big\}. \label{a19}
  \ee
 On the basis of \eqref{a10} and \eqref{a11},  we aim to obtain $(c_1, c_2)$ from the equations:
 \ber
 \re^{c_1}&=&\frac{ R_1(w_1, w_2,  \re^{c_2})\pm\sqrt{[R_1(w_1, w_2,  \re^{c_2})]^2-\frac{16\pi(ad-bc)(dN_1+bN_2) }{\lambda a(b+d)^2}\ito\re^{2u^1_0+2w_1}\ud x}}{2\ito\re^{2u^1_0+2w_1}}
 \nm\\&\equiv& g_1^{\pm}(\re^{c_2}), \label{a20}\\
  \re^{c_2}&=&\frac{ R_2(w_1, w_2,  \re^{c_1})\pm\sqrt{[R_2(w_1, w_2,  \re^{c_1})]^2-\frac{16\pi(ad-bc)(cN_1+aN_2) }{\lambda d(a+c)^2}\ito\re^{2u^2_0+2w_2}\ud x}}{2\ito\re^{2u^2_0+2w_2}}
 \nm\\&\equiv& g_2^{\pm}(\re^{c_1}).\label{a21}
 \eer

 To this end, we follow \cite{nota} and set
 \berr
 F^+(X)\equiv X-g_1^+(g_2^+(X)),\quad F^-(X)\equiv  X-g_1^-(g_2^-(X)),\\
 F^\pm(X)\equiv X-g_1^+(g_2^-(X)),\quad F^\mp(X)\equiv X-g_1^-(g_2^+(X)),
 \eerr
 so that the solutions of \eqref{a20} and \eqref{a21},  with all possible choices of signs: $ *=+, -, \pm, \mp$ corresponds  to the zeros of the function:
 \[F^*: [0, +\infty)\mapsto \mathbb{R}. \]
  At this point, as in \cite{nota}, it suffices to check  the following claims.

{\bf Claim 1. }  The functions $g_i^\pm(X)$ is strictly monotonic
with respect to $X>0$, $i=1, 2$.

 In fact, by  direct computation we have:
 \ber
  \frac{\ud g_1^\pm(X)}{\ud X}&=&\pm g_1^\pm(X)\frac{\frac{b(a+c)}{a(b+d)}\ito\re^{u^1_0+u_0^2+w_1+w_2}\ud x}
  {\sqrt{[R_1(w_1, w_2, X)]^2-\frac{16\pi(ad-bc)(dN_1+bN_2) }{\lambda a(b+d)^2}\ito\re^{2u^1_0+2w_1}\ud x}}, \label{a26}\\
  \frac{\ud g_2^\pm(X)}{\ud X}&=&\pm g_2^\pm(X)\frac{\frac{c(b+d)}{d(a+c)}\ito\re^{u^1_0+u_0^2+w_1+w_2}\ud x}
  {\sqrt{[R_2(w_1, w_2, X)]^2-\frac{16\pi(ad-bc)(cN_1+aN_2) }{\lambda d(a+c)^2}\ito\re^{2u^2_0+2w_2}\ud x}};  \label{a27}
 \eer
 and by definition (see \eqref{a20} and \eqref{a21})
 \be
  g_i^\pm(X)>0, \quad \forall\, X>0, \, i=1,2.\label{a28}
 \ee

 {\bf Claim 2. } For any $(w_1,w_2)\in \mathcal{A}$,
 there exits a unique $X^*(w_1,w_2)>0$ such that $F^*(X^*(w_1,w_2))=0$; with $*=+, -, \pm, \mp$.

 To  prove Claim 2,  observe that  $F^*(0)<0$,  with $ *=+, -, \pm, \mp$.  Next, we    check  easily that,
    \berr \lim\limits_{X\to+\infty}g^-_i(X)&=&0, \quad i=1, 2,\\
    \lim\limits_{X\to+\infty}\frac{g_1^+(X)}{X}&=&\frac{b(a+c)}{a(b+d)}\frac{\ito\re^{u^1_0+u_0^2+w_1+w_2}\ud x}{\ito\re^{2u^1_0+2w_1}\ud x}, \\
      \lim\limits_{X\to+\infty}\frac{g_2^+(X)}{X}&=&\frac{c(b+d)}{d(a+c)}\frac{\ito\re^{u^1_0+u_0^2+w_1+w_2}\ud x}{\ito\re^{2u^2_0+2w_2}\ud x},
  \eerr
 and consequently:
   \berr
  \lim\limits_{X\to+\infty}\frac{F^+(X)}{X}&=&1-\frac{bc}{ad}\frac{\left(\ito\re^{u^1_0+u_0^2+w_1+w_2}\ud x\right)^2}{\ito\re^{2u^1_0+2w_1}\ud x\ito\re^{2u^2_0+2w_2}\ud x}
  \ge\frac{ad-bc}{ad}>0,\\
 \lim\limits_{X\to+\infty}\frac{F^*(X)}{X}&=&1, \quad *=-, \pm, \mp.
   \eerr
 In particular,
 \[\lim\limits_{X\to+\infty}F^*(X)=+\infty, \quad *=+, -, \pm, \mp,\]
and from \eqref{a26} and \eqref{a27}, we see that
 \[\frac{\ud F^*(X)}{\ud X}>0, \quad *=\pm, \mp,\]
hence we deduce the statement of Claim 2  for  $*=\pm, \mp$.

 On the other hand, from \eqref{a26}-\eqref{a27} and \eqref{a17}-\eqref{a18} we obtain
 \berr
\frac{\ud F^+(X)}{\ud X}
  &=&1-\frac{b^2(a+c)^2}{a^2(b+d)^2}\frac{g_1^+(g_2^+(X))g_2^+(X)\left(\ito\re^{u^1_0+u_0^2+w_1+w_2}\ud x\right)^2}
  {\sqrt{[R_1(w_1,w_2, g_2^+(X))]^2 -\frac{16\pi(ad-bc)(dN_1+bN_2) }{\lambda a(b+d)^2}\ito\re^{2u^1_0+2w_1}\ud x}}\\
  &&\times\frac{1}{\sqrt{[R_2(w_1,w_2, X)]^2- \frac{16\pi(ad-bc)(cN_1+aN_2) }{\lambda d(a+c)^2}\ito\re^{2u^2_0+2w_2}\ud x}}
\\
  &>&1-\frac{g_1^+(g_2^+(X))}{X}=\frac{F^+(X)}{X}.
 \eerr
Similarly,  for $*=-$, we have:
  \berr
  \frac{\ud F^-(X)}{\ud X}>1-\frac{g_1^-(g_2^-(X))}{X}=\frac{F^-(X)}{X}.
 \eerr
 Thus, for $X>0$ the function   $\frac{F^*(X)}{X}$ is strictly increasing,  with $*=+, -$, and Claim 2 follows in this case as well. $\square$\\[3mm]

From the above discussion we see that, for any
$(w_1,w_2)\in\mathcal{A}$,  there exists a unique $c_j^*=c_j^*(w_1,
w_2)$ for $j=1, 2$  and $*=+, -, \pm, \mp$ such that
 \[ v_1^*=w_1+c_1^*(w_1,w_2), \quad  v_2^*=w_2+c_2^*(w_1,w_2), \quad *=+, -, \pm, \mp\]
satisfy \eqref{a8}-\eqref{a9}. Notice also that, by the above
property, $c_1^*$ and $c_2^*$ depend smoothly on
$(w_1,w_2)\in\mathcal{A}$.

We shall use those properties only for $*=+$, although it is
reasonable to expect that other choices  may lead to stronger
multiplicity results, as in \cite{nota}.

Thus, in what follows we   consider the functional
 \berr
 J^+_{\lambda}(w_1, w_2)=I_{\lambda}(w_1+c_1^+(w_1, w_2), w_2+c_2^+(w_1, w_2)), \quad (w_1, w_2)\in \mathcal{A}.
 \eerr

From \eqref{a7} and \eqref{a9'}, we see that
 \ber
  J^+_{ \lambda}(w_1, w_2)&=&\frac{d}{2b}\|\nabla w_1\|_2^2+\frac{a}{2c}\|\nabla w_2\|_2^2 +\ito\nabla w_1\cdot\nabla w_2\ud x \nm\\
 &&+\frac{\lambda}{2}\left[\left(1+\frac db\right)\ito\big(1-\re^{c_1^+}\re^{u^1_0+w_1}\big)\ud x+
 \left(1+\frac ac\right)\ito\big(1-\re^{c_2^+}\re^{u_0^2+w_2}\big)\ud x\right] \nm\\
 &&+\alpha_1c_1^+ +\alpha_2c_2^+-\frac{\alpha_1+\alpha_2}{2}\label{a29}
 \eer
with $\alpha_1, \alpha_2$ defined in \eqref{a7'}.

It is easy to check that the functional  $J^+_{\lambda}$ is
Frech\'{e}t differentiable in the interior of $\mathcal{A}$.
Moreover, if $(w_1, w_2)$ is a  critical point of $J^+_{\lambda}$
and lies in the interior of  $\mathcal{A}$,  then $ (w_1+c_1^+( w_1,
w_2), w_2+c_2^+(w_1, w_2))$ gives a critical point of
$I_{\lambda}$.

In the sequel, we   show that $J^+_{\lambda}$ is bounded from  below
and admits an interior minimum.

\begin{lemma}\label{lem1}
 For any $(w_1, w_2)\in \mathcal{A}$,  there holds:
  \be
  \re^{c_i^+}\ito \re^{u^i_0+w_i}\ud x\le |\Omega|,  \quad i=1, 2.\label{a30}
  \ee
\end{lemma}
\begin{remark}\label{rmk1}
 Using   Jensen's inequality, from \eqref{a30} follows that
  \[\re^{c_i^+}\le 1, \quad i= 1,2.\]
\end{remark}

{\bf Proof.}  \quad Using \eqref{a20}-\eqref{a21}, we have
 \ber
  \re^{c_1^+}&\le& \frac{\frac{ad-bc}{a(b+d)}\ito\re^{u_0^1+w_1}\ud x+\frac{b(a+c)}{a(b+d)}\re^{c^+_2}\ito\re^{u_0^1+u_0^2+w_1+w_2}\ud  x}{\ito\re^{2u_0^1+2w_1}\ud x},\label{a31}\\
    \re^{c_2^+}&\le& \frac{\frac{ad-bc}{d(a+c)}\ito\re^{u_0^2+w_2}\ud x+\frac{c(b+d)}{d(a+c)}\re^{c^+_1}\ito\re^{u_0^1+u_0^2+w_1+w_2}\ud x}{\ito\re^{2u_0^2+2w_2}\ud x}; \label{a32}
 \eer
 and so,  by using  \eqref{a31}-\eqref{a32} and  H\"{o}lder
inequality, we find
 \berr
  \re^{c_1^+}&\le& \frac{\frac{ad-bc}{a(b+d)}\ito\re^{u^1_0+w_1}\ud x}{\ito \re^{2u^1_0+2w_1}\ud x}
   +\frac{\frac{b(ad-bc)}{ad(b+d)}\ito\re^{u_0^2+w_2}\ud x\ito\re^{u_0^1+u_0^2+w_1+w_2}\ud  x}{\ito\re^{2u_0^1+2w_1}\ud x\ito\re^{2u_0^2+2w_2}\ud x}\nm\\
    &&+\frac{\frac{bc}{ad}\left(\ito\re^{u_0^1+u_0^2+w_1+w_2}\ud  x\right)^2\re^{c_1^+}}{\ito\re^{2u_0^1+2w_1}\ud x\ito\re^{2u_0^2+2w_2}\ud x}\nm\\
    &\le& \frac{\frac{ad-bc}{a(b+d)}\ito\re^{u^1_0+w_1}\ud x}{\ito \re^{2u^1_0+2w_1}\ud x}
   +\frac{\frac{b(ad-bc)}{ad(b+d)}\ito\re^{u_0^2+w_2}\ud x\ito\re^{u_0^1+u_0^2+w_1+w_2}\ud  x}{\ito\re^{2u_0^1+2w_1}\ud x\ito\re^{2u_0^2+2w_2}\ud x}+\frac{bc}{ad}\re^{c_1^+}.
 \eerr
Consequently,
 \ber
 \re^{c_1^+}\le \frac{\frac{d}{b+d}\ito\re^{u^1_0+w_1}\ud x}{\ito \re^{2u^1_0+2w_1}\ud x}
  +\frac{\frac{b}{b+d}\ito\re^{u_0^2+w_2}\ud x\ito\re^{u_0^1+u_0^2+w_1+w_2}\ud  x}{\ito\re^{2u_0^1+2w_1}\ud x\ito\re^{2u_0^2+2w_2}\ud x}. \label{a33}
 \eer
 Similarly, we obtain
 \ber
 \re^{c_2^+}\le \frac{\frac{a}{a+c}\ito\re^{u^2_0+w_2}\ud x}{\ito \re^{2u^2_0+2w_2}\ud x}
   +\frac{\frac{c}{a+c}\ito\re^{u_0^1+w_1}\ud x\ito\re^{u_0^1+u_0^2+w_1+w_2}\ud  x}{\ito\re^{2u_0^1+2w_1}\ud x\ito\re^{2u_0^2+2w_2}\ud x}. \label{a34}
 \eer
 Next, we can use  \eqref{a33}-\eqref{a34} and   H\"{o}lder inequality  to deduce that
 \berr
 \re^{c_1^+}\ito\re^{u^1_0+w_1}\ud x&\le& \frac{\frac{d}{b+d}\left(\ito\re^{u^1_0+w_1}\ud x\right)^2}{\ito \re^{2u^1_0+2w_1}\ud x}
   +\frac{\frac{b}{b+d}\ito\re^{u^1_0+w_1}\ud x\ito\re^{u_0^2+w_2}\ud x\ito\re^{u_0^1+u_0^2+w_1+w_2}\ud  x}{\ito\re^{2u_0^1+2w_1}\ud x\ito\re^{2u_0^2+2w_2}\ud x} \\
   &\le&|\Omega|, \\
   \re^{c_2^+}\ito\re^{u^2_0+w_2}\ud x&\le& \frac{\frac{a}{a+c}\left(\ito\re^{u^2_0+w_2}\ud x\right)^2}{\ito \re^{2u^2_0+2w_2}\ud x}
   +\frac{\frac{c}{a+c}\ito\re^{u_0^1+w_1}\ud x\ito\re^{u^2_0+w_2}\ud x\ito\re^{u_0^1+u_0^2+w_1+w_2}\ud  x}{\ito\re^{2u_0^1+2w_1}\ud x\ito\re^{2u_0^2+2w_2}\ud x}\\
     &\le&|\Omega|,
 \eerr
 and \eqref{a30}  is established. $\square$

The following property of functions in $\mathcal{A}$ was pointed out
first in \cite{nota2} and used in \cite{nota}. In our context, it
takes the following form:
 \begin{lemma}\label{lem2}
  For any $(w_1, w_2)\in \mathcal{A}$ and $s\in (0, 1)$,  we have
    \ber
   \ito\re^{u^1_0+w_1}\ud x&\le& \left(\frac{\lambda(ad-bc)}{16\pi a(dN_1+bN_2)}\right)^{\frac{1-s}{s}}
   \left(\ito\re^{su^1_0+sw_1}\ud x\right)^{\frac1s}, \label{a35}\\
     \ito\re^{u^2_0+w_2}\ud x&\le& \left(\frac{\lambda(ad-bc)}{16\pi d(cN_1+aN_2)}\right)^{\frac{1-s}{s}}
   \left(\ito\re^{su^2_0+sw_2}\ud x\right)^{\frac1s}. \label{a36}
    \eer
 \end{lemma}
{\bf Proof.}  \quad Although the proof of \eqref{a35}, \eqref{a36}
follows exactly  as in \cite{nota2, nota}, here  we give the proof
for completeness.
   Let $s\in (0, 1)$, $\gamma=\frac{1}{2-s}$ such that $s\gamma+2(1-\gamma)=1$. Then  using  H\"{o}lder inequality  and \eqref{a17} we have
   \berr
   \ito\re^{u^1_0+w_1}\ud x&\le&
   \left(\ito\re^{su^1_0+sw_1}\ud x\right)^\gamma\left(\ito\re^{2u^1_0+2w_1}\ud x\right)^{1-\gamma}\\
   &\le& \left(\frac{\lambda(ad-bc)}{16\pi a(dN_1+bN_2)}\right)^{1-\gamma}\left(\ito\re^{su^1_0+sw_1}\ud x\right)^\gamma\left(\ito\re^{u^1_0+w_1}\ud x\right)^{2(1-\gamma)},
   \eerr
which implies
 \berr
   \ito\re^{u^1_0+w_1}\ud x
   &\le& \left(\frac{\lambda(ad-bc)}{16\pi a(dN_1+bN_2)}\right)^{\frac{1-\gamma}{2\gamma-1}}\left(\ito\re^{su^1_0+sw_1}\ud x\right)^{\frac{\gamma}{2\gamma-1}}\\
 &=&\left(\frac{\lambda(ad-bc)}{16\pi a(dN_1+bN_2)}\right)^{\frac{1-s}{s}} \left(\ito\re^{su^1_0+sw_1}\ud x\right)^{\frac1s}.
   \eerr
Analogously, using  H\"{o}lder inequality  and \eqref{a18}, we can
get \eqref{a36}.  $\square$

Lemma \ref{lem2} will allow us to  show that the functional
$J^+_{\lambda}$ is coercive on $ \mathcal{A}$.  To  this purpose, we
need the following  well-known Moser--Trudinger inequality
(see\cite{aubi}):
 \be
 \ito \re^{w}\ud x \le C_1\exp\left(\frac{1}{16\pi}\|\nabla w\|_2^2\right), \quad \forall\, w\in H(\Omega): \ito w\ud x=0,\label{a37}
 \ee
 where $C_1$ is a positive constant depending on $\Omega$ only.
 \begin{lemma}\label{lem3}
    There exist suitable  constants $C_2>0$ and $C_3>0$ independent of $\lambda$ such that,  for every $(w_1, w_2)\in\mathcal{A}$ there holds:
    \ber
      J^+_\lambda(w_1, w_2) \ge C_2\left(\|\nabla w_1\|_2^2+\|\nabla w_2\|_2^2\right)
      -C_3(\ln\lambda+1).\label{a38}
    \eer
 \end{lemma}
 {\bf Proof.}\quad
 From \eqref{a20}-\eqref{a21}, we see that:
  \[
  \re^{c_1^+}\ge \frac{\frac{ad-bc}{2a(b+d)}\ito\re^{u^1_0+w_1}\ud x}{\ito \re^{2u^1_0+2w_1}\ud x},
   \quad \re^{c_2^+}\ge \frac{\frac{ad-bc}{2d(a+c)}\ito\re^{u_0^2+w_2}\ud x}{\ito \re^{2u^2_0+2w_2}\ud x}.
  \]
Thus,  by  the constraints \eqref{a17}-\eqref{a18}, we find:
 \berr
 \re^{c_1^+}\ge\frac{8\pi (dN_1+bN_2)}{\lambda(b+d)\ito\re^{u^1_0+w_1}\ud x}, \quad
   \re^{c_2^+}\ge\frac{8\pi(cN_1+aN_2)}{\lambda(a+c)\ito\re^{u^2_0+w_2}\ud x},
 \eerr
that is,
 \ber
  c_1^+&\ge& \ln\frac{8\pi(bN_1+dN_2)}{b+d}-\ln\lambda-\ln\ito\re^{u^1_0+w_1}\ud x,\label{a40}\\
  c_2^+&\ge& \ln\frac{8\pi(cN_1+aN_2)}{a+c}-\ln\lambda-\ln\ito\re^{u^2_0+w_2}\ud x.\label{a41}
 \eer

For any $s\in(0, 1)$,  using  Lemma \ref{lem2} and
Moser--Trudinger inequality \eqref{a37}, we have
 \ber
  &&\ln\ito\re^{u^1_0+w_1}\ud x\nm\\
  &&\le\frac{1-s}{s}\left[\ln\lambda+\ln\frac{ad-bc}{16\pi a(dN_1+bN_2)}\right]+\frac1s\ln\ito\re^{su^1_0+sw_1}\ud x\nm\\
   &&\le \frac{s}{16\pi}\|\nabla w_1\|_2^2+\frac{1-s}{s}\left[\ln\lambda+\ln\frac{ad-bc}{16\pi a(dN_1+bN_2)}\right]+\max\limits_{x\in\Omega}u^1_0+\frac1s\ln
   C_1; \label{a42}
  \eer
  and similarly:
  \ber
     \ln\ito\re^{u^2_0+w_2}\ud x
   \le \frac{s}{16\pi}\|\nabla w_2\|_2^2+\frac{1-s}{s}\left[\ln\lambda+\frac{ad-bc}{16\pi d(cN_1+aN_2)}\right]+\max\limits_{x\in\Omega}u^2_0+\frac1s\ln C_1.\label{a43}
 \eer
 Therefore from \eqref{a29}, \eqref{a9'}, and \eqref{a30},  for any given $\vep>0$, we obtain:
 \ber
  J^+_\lambda(w_1, w_2)&\ge& \left(\frac{d}{2b}-\frac\vep2\right)\|\nabla w_1\|_2^2
  +\left(\frac{a}{2c}-\frac{1}{2\vep}\right)\|\nabla w_2\|_2^2+\alpha_1c_1^+ + \alpha_2c_2^+,  \label{aa42}
 \eer
 where $\alpha_1$ and $\alpha_2$ are given in \eqref{a7'}. So,  with the optimal choice
 \[\vep=\frac12\left(\frac{d}{b}+\frac{c}{a}\right),\]
 we deduce:
 \ber
  J^+_\lambda(w_1, w_2)&\ge& \frac{ad-bc}{4ab}\|\nabla w_1\|_2^2
  +\frac{a(ad-bc)}{2c(ad+bc)}\|\nabla w_2\|_2^2+\alpha_1 c_1^+ + \alpha_2c_2^+.\label{aa43}
 \eer

 Then, from \eqref{aa43} and \eqref{a40}-\eqref{a43} for $s\in (0, 1)$ we conclude
  \ber
   J^+_\lambda(w_1, w_2)
   &\ge&\left(\frac{ad-bc}{4a}-\frac{s\alpha_1}{16\pi}\right)\|\nabla w_1\|_2^2+ \left(\frac{a(ad-bc)}{2c(ad+bc)}-\frac{s\alpha_2}{16\pi}\right)\|\nabla w_2\|_2^2\nm\\
   &&-\frac{\alpha_1+\alpha_2}{s}\ln\lambda-C, \label{a44}
  \eer
  with  $C$  a positive constant independent of $\lambda$ and $\alpha_1$ and $\alpha_2$ given in \eqref{a7'}.
  At this point, by choosing $s>0$ sufficiently small, the statement of  Lemma \ref{lem3} follows.  $\square$\\[3mm]

Since  $ J^+_\lambda(w_1,w_2)$ is weakly lower semicontinuous in
$\mathcal{A}$, by lemma \ref{lem3} we conclude that
$J^+_\lambda(w_1,w_2)$   attains the infimum  in $\mathcal{A}$.

 Next  we show that,  for    $\lambda$ is sufficiently large,   the minimizer of $J^+_\lambda$  belongs to the  interior  of $\mathcal{A}$.
 To this end, we observe the following:
 \begin{lemma}\label{lem4}
  There exists a positive constant $C_4$, independent of $\lambda$, such that,
   \be
   \inf\limits_{(w_1, w_2)\in \partial\mathcal{A}}J^+_\lambda(w_1, w_2)
    \ge  \frac{|\Omega|}{2}\min\left\{1+\frac db, 1+\frac ac\right\}\lambda-C_4(\ln\lambda+\sqrt{\lambda}+1).\label{a45}
   \ee
 \end{lemma}

{\bf Proof.} \quad  On the boundary of $\mathcal{A}$, we have
  \ber
  \left(\ito\re^{u^1_0+w_1}\ud x\right)^2&=&\frac{16\pi a(dN_1+bN_2)}{\lambda(ad-bc)}\ito\re^{2u^1_0+2w_1}\ud x\label{a46}
  \eer
   or
   \ber
    \left(\ito\re^{u^2_0+w_2}\ud x\right)^2&=&\frac{16\pi d(cN_1+aN_2)}{\lambda(ad-bc)}\ito\re^{2u^2_0+2w_2}\ud x.\label{a47}
   \eer

Suppose for example that  \eqref{a46} holds.  Then  using
\eqref{a33} and  H\"{o}lder inequality we obtain
 \berr
 \re^{c_1^+}\ito\re^{u^1_0+w_1}\ud x&\le& \frac{\frac{d}{b+d}\left(\ito\re^{u^1_0+w_1}\ud x\right)^2}{\ito \re^{2u^1_0+2w_1}\ud x}
   +\frac{\frac{b}{b+d}\ito\re^{u^1_0+w_1}\ud x\ito\re^{u_0^2+w_2}\ud x\ito\re^{u_0^1+u_0^2+w_1+w_2}\ud  x}{\ito\re^{2u_0^1+2w_1}\ud x\ito\re^{2u_0^2+2w_2}\ud x}\\
     &\le& \frac{16\pi ad(dN_1+bN_2)}{\lambda(b+d)(ad-bc)}+\frac{4b\sqrt{\pi a(dN_1+bN_2)|\Omega|}}{\sqrt{\lambda(ad-bc)}(b+d)}
  \eerr
  which  implies:
   \berr
   &&\frac{\lambda}{2}\left[\left(1+\frac db\right)\ito\big(1-\re^{c_1^+}\re^{u^1_0+w_1}\big)\ud x+\left(1+\frac ac\right)\ito\big(1-\re^{c_2^+}\re^{u_0^2+w_2}\big)\ud x\right]
   \\
   &&\ge\frac{|\Omega|}{2}\left(1+\frac db\right)\lambda-C_5\sqrt{\lambda}-C_6
   \eerr
with $C_5>0$ and $C_6>0$ suitable constants independent of
$\lambda$.

Now, estimating $c_1^+, c_2^+$ as in Lemma \ref{lem3}, we   arrive
at the estimate \eqref{a45}. $\square$

 At this point, we need to test $J^+_\lambda$ over a suitable  function in the interior of $\mathcal{A}$, for which the opposite inequality in \eqref{a45} holds.
  To  this end, we follow  \cite{nota} and recall that, for $\mu>0$ sufficiently large, there exist periodic solutions $v^i_{\mu}$ $(i=1, 2)$ for the problem:
   \ber
   \Delta v=\mu\re^{u^i_0+v}\big(\re^{u^i_0+v}-1\big)+\frac{4\pi N_i}{|\Omega|}\quad \text{in}\quad \Omega \label{a48}
   \eer
  such that $u^i_0+v^i_\mu<0$ in $\Omega$,   $c^i_\mu:=\frac{1}{|\Omega|}\ito v^i_\mu\ud  x\to 0$ and $w^i_\mu:=v^i_\mu-c^i_\mu\to -u^i_0$ pointwise  a.e. as
  $\mu\to +\infty$.  Those facts were  proved  in \cite{taran96}.

Since  $\re^{u^i_0}\in L^{\infty}(\Omega)$ $(i=1, 2)$, by the
dominated  convergence theorem, we have
  \[\re^{u^i_0+w^i_\mu}\to1 \quad \text{strongly in }\quad L^p(\Omega) \quad \text{for any } p\ge 1 \]
as $\mu\to +\infty$. In particular,
  \[\ito\re^{2u^i_0+2w^i_\mu}\ud x\to |\Omega|, \quad i=1,2  \]
 as $\mu\to +\infty$. Therefore, for $\lambda_0$ large and for fixed  $\vep\in (0, 1)$, we can find $\mu_\vep\gg1$, so  that $(w^1_{\mu_\vep},
 w^2_{\mu_\vep})$ lies in the interior of $\mathcal{A}$ for every $\lambda>\lambda_0$,  and the following holds:
   \ber
  \frac{\frac{(ad-bc)|\Omega|}{a+c}\left[a\ito\re^{2u^1_0+2w^1_{\mu_\vep}}\ud x+c|\Omega|\right]}
     {ad\ito\re^{2u^1_0+2w^1_{\mu_\vep}}\ud x\ito\re^{2u^2_0+2w^2_{\mu_\vep}}\ud  x-bc|\Omega|^2}\ge  1-\vep, \label{a49}\\
       \frac{\frac{(ad-bc)|\Omega|}{b+d}\left[d\ito\re^{2u^2_0+2w^2_{\mu_\vep}}\ud  x+b|\Omega|\right]}
     {ad\ito\re^{2u^1_0+2w^1_{\mu_\vep}}\ud x\ito\re^{2u^2_0+2w^2_{\mu_\vep}}\ud  x-bc|\Omega|^2}\ge1-\vep . \label{a50}
   \eer

  Using  Jensen's inequality,  and Remark \ref{rmk1}, in view of \eqref{a20}-\eqref{a21} by a straightforward calculation  we   get
  \ber
   \re^{c^+_1(w^1_{\mu_\vep}, w^2_{\mu_\vep})}&\ge& \frac{\frac{ad-bc}{a(b+d)}\ito\re^{u^1_0+w^1_{\mu_\vep}}\ud x+\frac{b(a+c)}{a(b+d)}\re^{c^+_2(w^1_{\mu_\vep}, w^2_{\mu_\vep})}\ito\re^{u^1_0+u^2_0+w^1_{\mu_\vep}
   +w^2_{\mu_\vep}}\ud x}{2\ito\re^{2u^1_0+2w^1_{\mu_\vep}}\ud x} \nm\\
   &&\times\left[1+\sqrt{1-\frac{16\pi a(dN_1+bN_2)}{\lambda(ad-bc)}\frac{\ito\re^{2u^1_0+2w^1_{\mu_\vep}}\ud x}{\left( \ito\re^{u^1_0+w^1_{\mu_\vep}}\ud x \right)^2}}\right] \nm\\
   &\ge&\frac{\frac{ad-bc}{a(b+d)}|\Omega|+\frac{b(a+c)}{a(b+d)}|\Omega|\re^{c^+_2(w^1_{\mu_\vep}, w^2_{\mu_\vep})}}{\ito\re^{2u^1_0+2w^1_{\mu_\vep}}\ud x}-\frac{8\pi a(dN_1+bN_2)}{\lambda(ad-bc)|\Omega|} \label{a51}
  \eer
 and similarly
  \ber
    \re^{c^+_2(w^1_{\mu_\vep}, w^2_{\mu_\vep})}&\ge&  \frac{\frac{ad-bc}{d(a+c)}|\Omega|+\frac{c(b+d)}{d(a+c)}|\Omega|\re^{c^+_1(w^1_{\mu_\vep}, w^2_{\mu_\vep})}}{\ito\re^{2u^2_0+2w^2_{\mu_\vep}}\ud x}
     -\frac{8\pi d(cN_1+aN_2)}{\lambda(ad-bc)|\Omega|}. \label{a52}
  \eer

Then inserting  \eqref{a52} into  \eqref{a51} we find,
  \berr
   \re^{c^+_1(w^1_{\mu_\vep}, w^2_{\mu_\vep})}
   &\ge&
    \frac{\frac{ad-bc}{a(b+d)}|\Omega|}{\ito\re^{2u^1_0+2w^1_{\mu_\vep}}\ud  x}-\frac{8\pi a(dN_1+bN_2)}{\lambda(ad-bc)|\Omega|}\\
    &&+\frac{\frac{b(a+c)|\Omega|}{a(b+d)}}{\ito\re^{2u^1_0+2w^1_{\mu_\vep}}\ud x}
    \left[ \frac{\frac{ad-bc}{d(a+c)}|\Omega|+\frac{c(b+d)}{d(a+c)}|\Omega|\re^{c^+_1(w^1_{\mu_\vep}, w^2_{\mu_\vep})}}{\ito\re^{2u^2_0+2w^2_{\mu_\vep}}\ud x}
     -\frac{8\pi d(cN_1+aN_2)}{\lambda(ad-bc)|\Omega|}\right]\\
    &\ge&\frac{\frac{ad-bc}{a(b+d)}|\Omega|}{\ito\re^{2u^1_0+2w^1_{\mu_\vep}}\ud x}+\frac{\frac{b(ad-bc)}{ad(b+d)}|\Omega|^2+\frac{bc}{ad}|\Omega|^2 \re^{c^+_1(w^1_{\mu_\vep}, w^2_{\mu_\vep})}}
    {\ito\re^{2u^1_0+2w^1_{\mu_\vep}}\ud x\ito\re^{2u^2_0+2w^2_{\mu_\vep}}\ud x}\nm\\
      &&-\frac{8\pi}{\lambda (ad-bc)|\Omega|}\left[a(dN_1+bN_2)+\frac{bd(a+c)(cN_1+aN_2)}{a(b+d)}\right],
  \eerr
which implies
  \ber
     \re^{c^+_1(w^1_{\mu_\vep}, w^2_{\mu_\vep})} &\ge&   \frac{\frac{(ad-bc)|\Omega|}{b+d}\left[d\ito\re^{2u^2_0+2w^2_{\mu_\vep}}\ud  x+b|\Omega|\right]}
     {ad\ito\re^{2u^1_0+2w^1_{\mu_\vep}}\ud x\ito\re^{2u^2_0+2w^2_{\mu_\vep}}\ud x-bc|\Omega|^2}\nm\\
      &&-\frac{8\pi ad}{\lambda (ad-bc)^2|\Omega|}\left[a(dN_1+bN_2)+\frac{bd(a+c)(cN_1+aN_2)}{a(b+d)}\right].\label{a53}
 \eer

Similarly, we get
  \ber
  \re^{c^+_2(w^1_{\mu_\vep}, w^2_{\mu_\vep})}&\ge&   \frac{\frac{(ad-bc)|\Omega|}{a+c}\left[a\ito\re^{2u^1_0+2w^1_{\mu_\vep}}\ud  x+c|\Omega|\right]}
     {ad\ito\re^{2u^1_0+2w^1_{\mu_\vep}}\ud x\ito\re^{2u^2_0+2w^2_{\mu_\vep}}\ud x-bc|\Omega|^2}\nm\\
     &&-\frac{8\pi ad}{\lambda (ad-bc)^2|\Omega|}\left[d(cN_1+aN_2)+\frac{ac(b+d)(dN_1+bN_2)}{d(a+c)}\right].\label{a54}
 \eer

Then,  by combining \eqref{a53}-\eqref{a54} and
\eqref{a49}-\eqref{a50}  we conclude that,
 \berr
 \re^{c^+_1(w^1_{\mu_\vep}, w^2_{\mu_\vep})} \ge 1-\vep-\frac{8\pi ad}{\lambda (ad-bc)^2|\Omega|}\left[a(dN_1+bN_2)+\frac{bd(a+c)(cN_1+aN_2)}{a(b+d)}\right],\\
 \re^{c^+_2(w^1_{\mu_\vep}, w^2_{\mu_\vep})}\ge 1-\vep-\frac{8\pi ad}{\lambda (ad-bc)^2|\Omega|}\left[d(cN_1+aN_2)+\frac{ac(b+d)(dN_1+bN_2)}{d(a+c)}\right];
 \eerr
 for all $\lambda>\lambda_0$.

 As a consequence, for all $\lambda>\lambda_0$,   we obtain that,
  \ber
   &&\ito\left(1-\re^{c^+_1(w^1_{\mu_\vep}, w^2_{\mu_\vep})}\re^{u^1_0+w^1_{\mu_\vep}}\right)\ud x\nm\\
   &&\le
    |\Omega|\vep+\frac{8\pi ad}{\lambda (ad-bc)^2}\left[a(dN_1+bN_2)+\frac{bd(a+c)(cN_1+aN_2)}{a(b+d)}\right],\label{a55}
    \\
    &&\ito\left(1-\re^{c^+_2(w^1_{\mu_\vep}, w^2_{\mu_\vep})}\re^{u^2_0+w^2_{\mu_\vep}}\right)\ud x\nm\\
    &&\le
  |\Omega|\vep+\frac{8\pi ad}{\lambda (ad-bc)^2}\left[d(cN_1+aN_2)+\frac{ac(b+d)(dN_1+bN_2)}{d(a+c)}\right]. \label{a56}
  \eer

\begin{lemma}\label{lem5}
     For $\lambda>0$ sufficiently large,  there holds:
    \be
    J^+_\lambda({w^1_{\mu_\vep}, w^2_{\mu_\vep})-  \inf\limits_{(w_1, w_2)\in \partial\mathcal{A}}J^+_\lambda}(w_1, w_2)
    <-1. \label{a57}
    \ee
\end{lemma}

 {\bf Proof.} \quad
     Using  \eqref{a55}-\eqref{a56} and the fact that   $c_1^+\le 0, \,c_2^+\le  0$, we conclude that, for any small $\vep\in (0, 1)$,  there exists a constant $C_\vep>0$ such that,
   \ber
   J^+_\lambda({w^1_{\mu_\vep}, w^2_{\mu_\vep})}
   &\le&   \frac{|\Omega|}{2}\left(2+\frac db+\frac ac\right)\vep\lambda+C_\vep.\label{a58}
   \eer

So  by  virtue of Lemma \ref{lem4} we  get
 \ber
 && J^+_\lambda({w^1_{\mu_\vep}, w^2_{\mu_\vep})}- \inf\limits_{(w_1, w_2)\in \partial\mathcal{A}}J^+_\lambda(w_1, w_2)
  \nm\\
  &&\le
  \frac{|\Omega|}{2}\min\left\{1+\frac db, 1+\frac ac\right\}\left[-1+\left(2+\frac db+\frac ac\right)\vep\right]\lambda+C(\ln\lambda+\sqrt{\lambda}+1),\label{a59}
 \eer
 with  $C>0$  independent of $\lambda$.    Clearly, \eqref{a57} easily follows from \eqref{a59} by choosing $\vep>0$ sufficiently small and  $\lambda>0$ sufficiently large. $\square$

 From Lemma \ref{lem3} and \ref{lem5}   we easily conclude:
  \begin{corollary}
   There exists $\bar{\lambda}>0$ such that for every $\lambda>\bar{\lambda}$,  the functional  $J^+_\lambda$ attains it minimum at a point
    $(w_{1,\lambda}, w_{2,\lambda})$, which lies  in the interior of $\mathcal{A}$.  Furthermore,
    \ber
     v^+_{1, \lambda}=w_{1,\lambda}+c_1^+(w_{1,\lambda}, w_{2,\lambda}), \quad  v^+_{2, \lambda}=w_{2,\lambda}+c_2^+(w_{1,\lambda}, w_{2,\lambda}) \label{a60}
    \eer
 defines a critical point for the functional $I_\lambda$ in $H(\Omega)\times H(\Omega)$,   namely a  (weak) solution  for  \eqref{a3}.
  \end{corollary}

Concerning such a solution we prove:
\begin{proposition}\label{pr2}
 Let $ (v^+_{1, \lambda}, v^+_{2, \lambda})$ be the solution of \eqref{a3} found above and defined by \eqref{a60}. We have

i)   \be\re^{u^i_0+v^+_{i, \lambda}}\to1, \quad \text{as}\quad
\lambda\to +\infty  \quad  (i=1, 2) \label{a61}\ee

    pointwise  a.e. in $\Omega$ and in  $L^p(\Omega)$ for any $p\ge1$.\\

ii)  $ (v^+_{1, \lambda}, v^+_{2, \lambda})$ defines a local minimum
for $I_\lambda$ in $H(\Omega)\times H(\Omega)$.
\end{proposition}

{\bf Proof.} \quad If we use \eqref{a38} together with \eqref{a58}
we readily find that,
  \ber
   \ito\left(\re^{u^i_0+v^+_{i, \lambda}}-1\right)^2\ud x\to 0, \quad \text{as}\quad \lambda\to +\infty, \quad i=1,2.
  \eer
 Since, by Proposition \ref{pr1}, we know that  $\re^{u^1_0+v^+_{1, \lambda}}<1,\,  \re^{u^2_0+v^+_{2, \lambda}}<1$   in $\Omega$,
 so by  the  dominated convergence  theorem we  conclude:
 \[ \re^{u^1_0+v^+_{1, \lambda}}\to1, \quad \re^{u^2_0+v^+_{2, \lambda}}\to1 \quad \text{as}\quad  \lambda\to+\infty\]
 pointwise a.e. in $\Omega$    and in  $L^p(\Omega), \, \forall\, p\ge1$ .

  To establish ii), we  check that for any $(w_1, w_2)\in \mathcal{A}$ and corresponding  $(c_1, c_2)$  given by  \eqref{a10},\eqref{a11}, we have:
   \berr
    \partial_{c_1}I_\lambda(w_1+c_1(w_1, w_2), w_2+c_2(w_1,w_2)) =0=
    \partial_{c_2}I_\lambda(w_1+c_1(w_1, w_2), w_2+c_2(w_1,w_2))
   \eerr
and
 \ber
    &&\partial^2_{c^2_1}I_\lambda(w_1+c_1(w_1, w_2), w_2+c_2(w_1,w_2))\nm\\
    &&= \frac{a(b+d)^2\lambda}{b(ad-bc)}\left[2\re^{2c_1}\ito\re^{2u_0^1+2w_1}\ud x-\re^{c_1}R_1(w_1, w_2, \re^{c_2})\right],\label{b59}\eer
    \ber
    &&\partial^2_{c^2_2}I_\lambda(w_1+c_1(w_1, w_2), w_2+c_2(w_1,w_2)) \nm\\
    &&= \frac{d(a+c)^2\lambda}{c(ad-bc)}\left[ 2\re^{2c_2}\ito\re^{2u_0^2+2w_2}\ud x-\re^{c_2}R_2(w_1, w_2, \re^{c_1})\right],\label{b60}\\
    &&\partial^2_{c_1c_2}I_\lambda(w_1+c_1(w_1, w_2), w_2+c_2(w_1,w_2)) \nm\\
    &&= -\frac{(a+c)(b+d)\lambda}{(ad-bc)}\re^{c_1}\re^{c_2} \ito\re^{u_0^1+u_0^2+w_1+w_2}\ud x.\label{b61}
\eer
 Next we use  \eqref{b59}-\eqref{b61} with $(c_1, c_2)=(c_1^+, c_2^+)$,  so that \eqref{a20} and \eqref{a21} hold with $+$ sign. Thus, for
  $v_i^+=w_i+c_i^+, \, i=1, 2$,  after straightforward calculation   we find:
 \berr
  &&\partial^2_{c^2_1}I_\lambda(v^+_1, v^+_2) \nm\\
  &&=\frac{a(b+d)^2\lambda}{b(ad-bc)}\left\{ \left[\frac{ad-bc}{a(b+d)}\ito\re^{u_0^1+v^+_1}\ud x+\frac{b(a+c)}{a(b+d)}\ito\re^{u_0^1+u_0^2+v^+_1+v^+_2}\ud  x\right]^2\right.\nm\\
  &&\left.\quad-\frac{16\pi(ad-bc)(dN_1+bN_2) }{\lambda a(b+d)^2}\ito\re^{2u^1_0+2v^+_1}\ud x\right\}^{\frac12},\\
  &&\partial^2_{c^2_2}I_\lambda(v^+_1, v^+_2)\nm\\
  &&=\frac{d(a+c)^2\lambda}{c(ad-bc)}\left\{\left[\frac{ad-bc}{d(a+c)}\ito\re^{u_0^2+v^+_2}\ud x+\frac{c(b+d)}{d(a+c)}\ito\re^{u_0^1+u_0^2+v^+_1+v^+_2}\ud x\right]^2\right.\nm\\
  &&\left.\quad-\frac{4\pi(ad-bc)(cN_1+aN_2) }{\lambda d(a+c)^2}\ito\re^{2u^2_0+2v^+_2}\ud x\right\}^{\frac12}.
 \eerr

In case $(w_1, w_2)$ lies in the interior of $\mathcal{A}$,  then we
can use the strict inequality in \eqref{a17}, \eqref{a18} and obtain
  \berr
   \partial^2_{c^2_1}I_\lambda(v^+_1, v^+_2)>\frac{(a+c)(b+d)\lambda}{(ad-bc)}\ito\re^{u_0^1+u_0^2+v^+_1+v^+_2}\ud  x,\\
   \partial^2_{c^2_2}I_\lambda(v^+_1, v^+_2)>\frac{(a+c)(b+d)\lambda}{(ad-bc)}\ito\re^{u_0^1+u_0^2+v^+_1+v^+_2}\ud  x.
  \eerr
Therefore we have checked that, if $(w_1, w_2)$ is an interior point
of  $ \mathcal{A}$ then the Hessian matrix of $I_\lambda(w_1+c_1,
w_2+c_2)$ with respect to $(c_1, c_2)$ is strictly positive definite
at $(c_1^+(w_1, w_2), c_2^+(w_1, w_2))$. We apply such property,
near the critical point $(v^+_{1, \lambda}, v^+_{2, \lambda})$.
Indeed, by continuity, for $\delta>0$ sufficiently small, we can
ensure that, if
 $(v_1, v_2)=(w_1+c_1, w_2+c_2)$  satisfies:
  \[ \|v_1-v^+_{1, \lambda}\|+\|v_2-v^+_{2, \lambda}\|\le \delta, \]
  then  $(w_1, w_2)$ belongs to the interior of $\mathcal{A}$  and
  \berr
   I_\lambda(v_1, v_2)= I_\lambda(w_1+c_1, w_2+c_2)&\ge& I_\lambda(w_1+c^+_1(w_1, w_2), w_2+c^+_2(w_1, w_2))\\
   &=&J^+_\lambda(w_1, w_2)\ge  I_\lambda(v^+_{1, \lambda}, v^+_{2, \lambda}).
  \eerr
Consequently, $(v^+_{1, \lambda}, v^+_{2, \lambda})$  defines a
local minimizer for $ I_\lambda$ in $H(\Omega)\times H(\Omega)$, as
desired. $\square$

\subsection{Mountain-Pass solution}

To complete the proof of Theorem \ref{thb1}, it remains to establish
the existence of a second solution. Again we use the variational
approach and show that the functional $I_\lambda$ admits also a
``saddle'' critical point of ``mountain-pass'' type. We start to
establish the following:
\begin{lemma}\label{lem9}
 The functional $I_\lambda$ satisfies the  (P.S.) condition in $H(\Omega)\times H(\Omega)$. Namely,   every sequence $(v_{1, n}, v_{2, n})\in H(\Omega)\times H(\Omega) $ satisfying:
   \ber
      I_\lambda(v_{1, n}, v_{2, n})\to m_0\quad   \text{as}\quad n\to +\infty, \label{a68}\\
      \|I'_\lambda(v_{1, n}, v_{2, n})\|_*\to 0\quad   \text{as}\quad n\to +\infty,\label{a69}
  \eer
admits a strongly convergent subsequence in  $H(\Omega)\times
H(\Omega)$, where $m_0$ is a constant and $\|\cdot\|_*$ denotes the
norm of dual space of $H(\Omega)\times H(\Omega)$.
\end{lemma}

{\bf Proof.} \quad Let $\vep_n=\|I'_\lambda(v_{1,n}, v_{2,n})\|_*\to
0, \,\,n\to +\infty$, and observe that $\forall \, (\psi_1,
\psi_2)\in H(\Omega)\times H(\Omega)$, we have:
 \ber
  &&I'_\lambda(v_{1, n}, v_{2, n})[(\psi_1, \psi_2)]\nm\\
  &&=\frac db\ito \nabla v_{1,n}\cdot\nabla\psi_1 \ud x +\frac ac\ito \nabla v_{2,n}\cdot\nabla\psi_2 \ud x+\ito \nabla v_{1,n}\cdot\nabla \psi_2\ud x+ \ito \nabla v_{2,n}\cdot\nabla \psi_1\ud x\nm\\
  &&\quad+\frac{\lambda}{ab(ad-bc)}\ito Q_1(v_{1,n}, v_{2, n})\left[a(b+d)\re^{u^0_1+v_{1,n}}\psi_1-b(a+c)\re^{u^0_2+v_{2,n}}\psi_2\right]\ud x \nm\\
  &&\quad+ \frac{\lambda(a+c)^2}{ac}\ito Q_2(v_{2,n})\re^{u_0^2+v_{2,n}}\psi_2\ud x+\frac{\alpha_1}{|\Omega|}\ito \psi_1\ud x+\frac{\alpha_2}{|\Omega|}\ito \psi_2\ud x\label{a70}
 \eer
and
  \be
   |I'_\lambda(v_{1,n}, v_{2,n})(\psi_1,\psi_2)|\le \vep_n(\|\psi_1\|+\|\psi_2\|).\label{a71}
  \ee

 In particular, if we take $\psi_1=\psi_2=\psi\in H(\Omega)$ in \eqref{a70} we find:
   \ber
   I'_\lambda(v_{1, n}, v_{2, n})[(\psi, \psi)]&=&\ito \nabla \left[\frac{b+d}{b}v_{1,n}+\frac{a+c}{c}v_{2,n}\right]\cdot\nabla\psi\nm\\
    &&+2\lambda\ito Q(v_{1,n}, v_{2,n})\psi \ud x +\frac{\lambda}{ab}\ito Q_1(v_{1,n}, v_{2,n})\psi\ud x\nm\\
    &&+\frac{\lambda(a+c)^2}{ac}\ito Q_2(v_{2,n})\psi\ud x+\frac{\alpha_1+\alpha_2}{|\Omega|}\ito\psi\ud x, \label{a72}
   \eer
where we recall that $Q, Q_1, Q_2$ and $\alpha_1, \alpha_2$ are
defined in \eqref{a7'}.

As a consequence for $\psi\equiv1$, we deduce that,
     \[ \ito Q_1^2(v_{1,n}, v_{2,n})\ud x+\ito Q_2^2(v_{2,n})\ud x\le C\]
 for some suitable constant $C>0$. In particular,
   \ber
   \ito \re^{2(u_0^1+v_{1,n})}\ud x+\ito \re^{2(u_0^2+v_{2,n})}\ud x\le C\label{a73}
   \eer
with  a (possible different) $C>0$.

Decompose $v_{j,n}=w_{j,n}+c_{j,n}$ with $\ito w_{j,n}\ud x=0, \,
c_{j,n}=\frac{1}{|\Omega|}\ito v_{j,n}\ud x\, (j=1, 2)$ and observe
that, (by assumption)
 \ber
 I_\lambda(v_{1,n},v_{2,n})&=&\frac12\left[\left(\frac db-\frac ca \right)\|\nabla v_{1,n}\|_2^2 +\left\|\nabla \left(\sqrt{\frac ca}v_{1,n}+\sqrt{\frac ac}v_{2,n}\right)\right\|_2^2\right]\nm\\
 &&+\lambda\ito Q(v_{1,n}, v_{2, n}) \ud x+\alpha_1c_{1,n}+\alpha_2c_{2,n} \to m_0 \quad   \text{as}\quad n\to +\infty.\label{a74}
 \eer
 Moreover, from \eqref{a73}  and Jensen's inequality, we find
    \be
      c_{j,n}\le c_0, \quad \forall\, n\in \mathbb{N}, \quad j=1, 2\label{a75}
    \ee
with suitable $c_0>0$.

 Next, let
   \be
     z_n=\frac{b+d}{b}w_{1,n}+\frac{a+c}{c}w_{2, n}\label{a76}
   \ee
 so that $\ito z_n\ud x=0$.  If we take in \eqref{a72} $\psi=z_n^+=\max\{z_n, 0\}$, from
 \eqref{a73} we find
  \ber
  && \|\nabla z_n^+\|_2^2+\frac{\lambda}{ad-bc}\ito \left[\sqrt{\frac ab}(b+d)\re^{u_0^1+v_{1,n}}-\sqrt{\frac dc}(a+c)\re^{u_0^2+v_{2,n}}\right]^2z_n^+\ud x\nm\\
  && +\frac{2\lambda(a+c)(b+d)}{ad-bc}\left(\sqrt{\frac{ad}{bc}}-1\right)\ito \re^{u_0^1+v_{1,n}}\re^{u_0^2+v_{2,n}}z_n^+\ud x\nm\\
  &&\le C(\|z_n^+\|_2+\vep_n\|z_n^+\|).\label{a77}
  \eer

Then from \eqref{a77} and Poincar\'{e} inequality we obtain
 \be
  \ito \re^{u_0^1+v_{1,n}}\re^{u_0^2+v_{2,n}}z_n^+\ud x\le C(\|\nabla w_{1,n}\|_2+\|\nabla w_{2, n}\|_2).\label{a78}
   \ee

To proceed further we choose $\psi_1=w_{1,n}$ and $\psi_2=w_{2,n}$
in \eqref{a70} and after straightforward calculations we obtain:
 \ber
  &&I'(v_{1,n}, v_{2,n})[(w_{1,n}, w_{2,n})]=\left(\frac db-\frac ca\right)\|\nabla w_{1, n}\|_2^2+\left\|\nabla \left(\sqrt{\frac ca}w_{1,n}+\sqrt{\frac ac}w_{2,n}\right)\right\|_2^2 \nm\\
  &&+\frac{\lambda}{ab(ad-bc)}\left[\ito a^2(b+d)^2\re^{2(u_0^1+v_{1,n})}w_{1,n}+b^2(a+c)^2\re^{2(u_0^2+v_{2,n})}w_{2,n}\right.\nm\\
  &&\left.-ab(a+c)(b+d)\re^{u_0^1+v_{1,n}}\re^{u_0^2+v_{2,n}}(w_{1,n}+w_{2,n})\right]+ \frac{\lambda(a+c)^2}{ac}\ito\re^{2(u_0^2+v_{2,n})}w_{2,n}\ud x\nm\\
  &&-\lambda\left[\frac{b+d}{b}\ito \re^{u_0^1+v_{1,n}}w_{1,n}\ud x+\frac{a+c}{c}\ito \re^{u_0^2+v_{2,n}}w_{2,n}\ud x\right].\label{a79}
 \eer
Clearly, in view of \eqref{a73} we can estimate
  \[\left|\ito \re^{u_0^j+v_{j,n}}w_{j,n}\ud x\right|\le C\|w_{j,n}\|_2.\]
 While from \eqref{a75} we get
   \berr
   \ito \re^{2(u_0^j+v_{j,n})}w_{j,n}\ud x&=&\ito \re^{2(u_0^j+c_{j,n})}\left(\re^{w_{j,n}}-1\right)w_{j,n}\ud x+\ito \re^{2(u_0^j+c_{j,n})}w_{j,n} \ud x\\
   &\ge&-\re^{c_0}\|\re^{2u_0^j}\|_2\|w_{j,n}\|_2\\
   &\ge&-C \|\nabla w_{j,n}\|_2, \quad j=1, 2.
   \eerr

Furthermore, we see that
   \berr
    &&\ito\re^{u_0^1+v_{1,n}}\re^{u_0^2+v_{2,n}}(w_{1,n}+w_{2,n})\ud x\\
    &&\le \ito\re^{u_0^1+v_{1,n}}\re^{u_0^2+v_{2,n}}(w_{1,n}+w_{2,n})_+\ud x\\
    && =\int\limits_{\{w_{1,n}\le0\le w_{2,n}\}} \re^{u_0^1+c_{1,n}}\left(\re^{w_{1,n}}-1\right)\re^{u_0^2+v_{2,n}}(w_{1,n}+w_{2,n})_+\ud x\\
    && \quad+\int\limits_{\{w_{1,n}\le0\le w_{2,n}\}}\re^{u_0^1+c_{1,n}}\re^{u_0^2+v_{2,n}}(w_{1,n}+w_{2,n})_+\ud x\\
    &&\quad +\int\limits_{\{w_{2,n}\le0\le w_{1,n}\}} \re^{u_0^2+c_{2,n}}\left(\re^{w_{2,n}}-1\right)\re^{u_0^1+v_{1,n}}(w_{1,n}+w_{2,n})_+\ud x\\
    &&  \quad+\int\limits_{\{w_{2,n}\le0\le w_{1,n}\}}\re^{u_0^2+c_{2,n}}\re^{u_0^1+v_{1,n}}(w_{1,n}+w_{2,n})_+\ud x\\
    &&\quad +\int\limits_{ \{w_{1,n}>0\}\cap\{w_{2,n}>0\}} \re^{u_0^1+v_{1,n}} \re^{u_0^2+v_{2,n}}(w_{1,n}+w_{2,n})_+\ud x\\
    &&\le C\left(\|\nabla w_{1,n}\|_2+\|\nabla w_{2,n}\|_2\right)+\left(\frac{b}{b+d}+\frac{c}{a+c}\right)\ito\re^{u_0^1+v_{1,n}} \re^{u_0^2+v_{2,n}}z_n^+\ud x.
   \eerr
 Thus from \eqref{a78}, we conclude:
   \berr
    \ito\re^{u_0^1+v_{1,n}} \re^{u_0^2+v_{2,n}}(w_{1,n}+w_{2,n})\ud x\le C\left(\|\nabla w_{1,n}\|_2+\|\nabla w_{2,n}\|_2\right).
   \eerr
Using  the estimates above, together with \eqref{a79} and
\eqref{a69}, we conclude that
  \ber
   \left(\frac db-\frac ca\right)\|\nabla w_{1,n}\|_2+\left\|\nabla \left(\sqrt{\frac ca}w_{1,n}+\sqrt{\frac ac}w_{2,n}\right)\right\|_2^2 \le C. \label{a80}
  \eer

 Now from \eqref{a74}, \eqref{a75} and  \eqref{a80}, we deduce that $\{c_{j,n}\}$
 is also uniformly bounded from below, for $j=1,2$.

 Consequently,  $\{v_{j,n}\}$ is a uniformly bounded sequence in $H(\Omega)$, for $j=1,2$. So, along a subsequence, (denoted the same
 way),  and for suitable $v_j\in H(\Omega)\,  (j=1,2)$ we have:
    \berr
     &&v_{j,n}\to v_j \, \text{as}\,\,  n\to+\infty \,\, \text{weakly in }\,\, H(\Omega),\,\text{ and strongly in}\,\,  L^p(\Omega), \, p\ge1\,\, \text{and pointwise a.e. in}\, \Omega;\\
     &&\re^{u_0^j+v_{j,n}}\to\re^{u_0^j+v_j} \,\, \text{as}\,\,  n\to+\infty \,\,\text{in}\,\,  L^p(\Omega), \, p\ge1;\,\,  j=1,2.
    \eerr

In particular, $(v_1, v_2)$ is a critical point for $I_\lambda$ and
by the above convergence properties we have:
   \berr
   &&\left(\frac db-\frac ca\right)\|\nabla (v_{1,n}-v_1)\|_2^2+\left\|\nabla \left(\sqrt{\frac ca}(v_{1,n}-v_1)+\sqrt{\frac ac}(v_{2,n}-v_2)\right)\right\|_2^2\\
   &&= \left(I_\lambda'(v_{1,n},v_{2,n})-I'_\lambda(v_1,v_2)\right)[(v_{1,n}-v_1, v_{2,n}-v_2)]+o(1)\to 0, \quad \text{as}\quad  n\to +\infty.
   \eerr
Thus,  $v_{j,n}\to v_j$ strongly in $H(\Omega)$ as   $n\to +\infty, \, j=1,2$; and the proof of Lemma \ref{lem9} is completed.  $\square$\\

To proceed further, we need to use the minimization property of
$(v^+_{1, \lambda}, v^+_{2, \lambda})$ as given in Proposition
\ref{pr2}-(ii). In case it  defines a degenerate (local) minimum for
$I_\lambda$, in the  sense  that for every $\delta>0$ sufficiently
small,
  \berr
  \inf\limits_{\left\{\|v_1-v_{1,\lambda}^+\|+\|v_2-v_{2,\lambda}^+\|=\delta\right\}} I_\lambda(v_1,v_2) =I_\lambda(v^+_{1, \lambda}, v^+_{2, \lambda}).
   \eerr
then we obtain a $1$-parameter family of (degenerate) local minima
of $I_\lambda$, (see Corollary 1.6 of  \cite{gho}), and  the
conclusion of Theorem \ref{thb1} is  obviously  established in this
case.

Hence, we suppose that $(v^+_{1, \lambda}, v^+_{2, \lambda})$
defines a strict local minimum for $I_\lambda$.  In particular,  for
$\delta>0$
 sufficiently small,  the following holds:
     \be
     I_\lambda(v_{1,\lambda}^+, v_{2,\lambda}^+)<\inf\limits_{\left\{\|v_1-v_{1,\lambda}^+\|+\|v_2-v_{2,\lambda}^+\|= \delta\right\}}  I_\lambda(v_1, v_2)
       :=\gamma_0. \label{b75}
     \ee

 In addition, we observe that,
  \[ I_\lambda(v^+_{1, \lambda}-\xi, v^+_{2, \lambda}-\xi)\to -\infty, \quad \text{as}\quad \xi\to +\infty.\]
 Hence,  for fixed $\bar{\xi}>1$ sufficiently large and  $\bar{v}_i=v_{i,\lambda}^+-\bar{\xi} , \,\,i=1,2, $ we find
   \ber
  \|v_{1,\lambda}^+-\bar{v}_1\|+\|v_{2,\lambda}^+-\bar{v}_2\|>\delta \quad \text{and}\quad I_\lambda(\bar{v}_1,\bar{v}_2)
    <I_\lambda(v_{1,\lambda}^+, v_{2,\lambda}^+). \label{b76}
   \eer

Lemma \ref{lem9} together with \eqref{b75} and \eqref{b76}, allow us
to use the ``mountain-pass'' lemma of  Ambrosetti-Rabinowitz
\cite{amra} and conclude the existence of a second critical point
$(\tilde{v}_{1,\lambda}, \tilde{v}_{2,\lambda})$ for $I_\lambda$
satisfying:
   \be
   I_{\lambda}(\tilde{v}_{1,\lambda}, \tilde{v}_{2,\lambda})\ge\gamma_0>I_{\lambda}(v_{1,\lambda}^+, v_{2,\lambda}^+).\label{a82}
  \ee
By virtue of \eqref{a82},  such critical point  yields to a solution
for \eqref{a3} distinct from $(v^+_{1, \lambda}, v^+_{2, \lambda})$.
This
completes the proof of Theorem \ref{thb1}.\\

 It would be interesting to see whether, as for the gauge group $SU(3)$, a stronger multiplicity result holds, in  relation  to each vacua state of the system.

 For example, it is natural to expect that the ``mountain-pass'' solution is asymptotically gauge equivalent to the unbroken  vacuum for $\lambda\to+\infty$;
 as it occurs in the Abelian case, see \cite{cho1}.\\

\noindent{\bf Acknowledgement}

The research of X. Han was supported by the by the National Natural Science Foundation of China under grant 11201118, Henan Basic Science and Frontier Technology Program Funds under grant
112300410054, and the Key Youth Teacher Foundation of Department Education of Henan Province under grant 2011GGJS-210.
The research of G. Tarantello has been supported by FIRB-Ideas project: Analysis and Beyond, and by PRIN-project: Nonlinear elliptic problems in the study of vortices and related topics.

\end{document}